\documentclass[aps, floatfix, nofootinbib, amsmath, amssymb, eqsecnum, twocolumn, fleqn, authoryear]{revtex4-1}
\usepackage[french]{babel}
\usepackage{color}
\usepackage{graphicx}
\usepackage{float}
\usepackage{ifthen}
\usepackage{bm}
\usepackage{upgreek}
\usepackage{hyperref}
\usepackage{soul}
\usepackage[LGR,T1]{fontenc}
\DeclareSymbolFont{upgreek}{LGR}{cmr}{m}{n}
\SetSymbolFont{upgreek}{bold}{LGR}{cmr}{bx}{n}
\usepackage{amsmath}	
\usepackage{amssymb}	

\setcitestyle{authoryear,round,sort}
\usepackage{rotating}
\usepackage{url}
\usepackage{bbold}
\usepackage{aas_macros}
\usepackage{verbatim}
\usepackage{algorithm}
\usepackage{algpseudocode}
\usepackage{enumitem}

\DeclareMathAlphabet{\mathpzc}{OT1}{pzc}{m}{it}

\newcommand{\mvec}[1]{\bm{#1}}

\newcommand{\myvec}[1]{\mvec{#1}}

\usepackage[dvipsnames]{xcolor}
\hypersetup{
  colorlinks   = true, 
  urlcolor     = RoyalBlue, 
  linkcolor    = RoyalBlue, 
  citecolor   = MidnightBlue 
}

\begin{document}
	\author{Doogesh Kodi Ramanah}
		\email{ramanah@iap.fr}
		\affiliation{Sorbonne Universit\'e, CNRS, UMR 7095, Institut d'Astrophysique de Paris, 98 bis boulevard Arago, 75014 Paris, France}
		\affiliation{Sorbonne Universit\'e, Institut Lagrange de Paris, 98 bis boulevard Arago, 75014 Paris, France}    
	\author{Tom Charnock}
		\email{charnock@iap.fr}
		\affiliation{Sorbonne Universit\'e, CNRS, UMR 7095, Institut d'Astrophysique de Paris, 98 bis boulevard Arago, 75014 Paris, France}    
	\author{Guilhem Lavaux}
		\email{lavaux@iap.fr}
		\affiliation{Sorbonne Universit\'e, CNRS, UMR 7095, Institut d'Astrophysique de Paris, 98 bis boulevard Arago, 75014 Paris, France}
			             
    \title{Painting halos from cosmic density fields of dark matter with physically motivated neural networks}
    
	\begin{abstract}
    We present a novel halo painting network that learns to map approximate 3D dark matter fields to realistic halo distributions.
    This map is provided via a physically motivated network with which we can learn the non-trivial local relation between dark matter density field and halo distributions without relying on a physical model.
    Unlike other generative or regressive models, a well motivated prior and simple physical principles allow us to train the mapping network quickly and with relatively little data.
    In learning to paint halo distributions from computationally cheap, analytical and non-linear density fields, we bypass the need for full particle mesh simulations and halo finding algorithms.
    Furthermore, by design, our halo painting network needs only local patches of dark matter density to predict the halos, and as such, it can predict the 3D halo distribution for any arbitrary simulation box size.
    Our neural network can be trained using small simulations and used to predict large halo distributions, as long as the resolutions are equivalent.
    We evaluate our model's ability to generate 3D halo count distributions which reproduce, to a high degree, summary statistics such as the power spectrum and bispectrum, of the input or reference realizations.
	\end{abstract}
    
	\maketitle

    \section{Introduction} \label{intro_WGAN}

    Investigating the formation and evolution of dark matter halos, as the key building blocks of cosmic large-scale structure, is essential for constraining various cosmological models and further understanding our Universe.
    The highly non-linear dynamics involved nevertheless renders this a complex problem, with $N$-body simulations currently the only tool to compute the non-linear gravitational evolution from initial conditions \citep[e.g.][]{springel2005gadget2}, yielding mock dark matter halo catalogues as the main output.
    The resulting catalogues of positions, velocities and masses of halos are necessary for cosmological inference from galaxy surveys.
    As an example, running very large simulations of pure dark matter, such as \textsc{fur-deus} \citep{Alimi2012}, to generate fake observations of the full Universe several times is not feasible, and requires a large amount of memory and disk storage.
    A way to emulate such simulations, quickly and reliably, would be of use to a wide community as a new method for data analysis and light cone production for the next cosmological survey missions such as Euclid \citep{euclid2011report} and Large Synoptic Survey Telescope (LSST) \citep{lsst2008summary}.
    
\medskip
    With the recent developments in the field of machine learning, deep generative modelling techniques have emerged as a viable tool to construct emulators of expensive simulations.
    In this work, we present such a deep learning approach to generate the 3D halo distribution from dark matter simulations.
    Using our construction, the neural network is used to learn the mapping from the dark matter density to halo fields and therefore predicts the abundance of halos at a given position based on the large-scale density distribution.
    Once trained, the emulator is capable of rapidly predicting simulations of halo distribution based on a non-linearly evolved density field.
    Furthermore, by learning this mapping for different halo mass bins, we can also predict the mass distribution of the halos.
    
\medskip
	A key aspect of our approach is that the neural network is able to \emph{paint} a halo count distribution from a numerically cheap non-linear density field, such as a realization obtained via Lagrangian Perturbation Theory (LPT), which requires negligible computational resources on modern machines relative to $N$-body simulations.
    The interest of this technique lies in the possibility that most of the cosmological dependence of the observed matter distribution in the $N$-body simulation is already encoded in the inexpensive LPT simulation.
    As a result, this approach would eliminate the need to run a full particle mesh simulation, thereby allowing detailed analyses of state-of-the-art surveys to be feasible on regular computing facilities. 
    The idea of painting complex astronomical objects has been implemented in the past, notably for galaxies with \textsc{molusc} \citep{sousbie2008molusc} and \textsc{lymas} \citep{peirani2014lymas}.
    Another related work is the \textsc{pinocchio} algorithm \citep{monaco2002pinocchioI, monaco2002pinocchioII} for identifying dark matter halos in a given numerical realization of the linear density field.
    However, our aim here is to build an automated model generator with even higher accuracy and insensitivity to the underlying cosmology.
    
\medskip    
    We take inspiration from a recently proposed variant of generative models, known as generative adversarial networks (GANs) \citep{goodfellow2014generative}, which have met considerable success with a range of applications, such as generating extremely realistic fake celebrities \citep{karras2017progressive, karras2018style} and artificial bedroom images \citep{radford2015unsupervised}.
    In particular, we will use the key ideas in training WGANs, i.e. GANs optimized using the Wasserstein distance \citep{arjovsky2017wgan}, to ensure that our network is able to paint halos well.
    The GANs, and variants thereof, are described in more depth in Section \ref{GANs}.
    
\medskip    
    Neural networks have recently been employed for various aspects of large-scale structure analysis.
    \citet{he2018learning} devised a deep neural network to predict the non-linear cosmic structure formation from linear perturbation theory, with the network architecture based on the U-Net \citep{ronneberger2015UNet} learning model.
    \citet{zhang2019darkmatter} constructed a two-phase convolutional neural network architecture to map 3D dark matter fields to the corresponding galaxy distribution in hydrodynamic simulations.
    \citet{berger2018volumetric} implemented a 3D deep convolutional neural network to generate mock halo catalogues by identifying protohalos directly from the cosmological initial conditions.
    This was preceded by a similar work by \citet{luciesmith2018machine}, where a random forest classifier was used to trace the halos formed in $N$-body simulations back to their initial conditions.
    \citet{modi2018cosmological} proposed a framework, with the end product being the converse, to reconstruct initial conditions from the halo fields using the multilayer perceptron, i.e. fully connected neural networks.
    Deep convolutional networks were also used to classify the distinct components of the cosmic web such as filaments and walls from $N$-body simulations \citep{aragon2018classifying}.
       
\medskip    
    The paper is organized as follows.
    We outline the underlying conceptual framework of the GANs in Section~\ref{GANs}, followed by a description of the recently proposed WGAN.
    Section~\ref{network_refinements} describes several other network refinements which we implement in designing our network architecture.
    Section~\ref{velmass} illustrates the relevant aspects of the dark matter simulations used in the training and validation of our neural network, as described in Section~\ref{implementation_WGAN}.
    We follow up by investigating the performance of the algorithm in terms of various diagnostics, as detailed in Section~\ref{results_WGAN}.
    Finally, in Section~\ref{conclusions_WGAN}, we summarize our main findings and discuss the areas of application where our halo painting network, tailored for mapping dark matter distributions to halo fields, can be optimized.
    
    \section{Generative adversarial networks} \label{GANs}

    Generative adversarial networks were first proposed in the seminal work of \citet{goodfellow2014generative}, and have emerged as powerful generative models, albeit with some limitations, as described below, which have been addressed by recent developments.
    In the standard GAN framework, generative modelling is formulated as a game between two competing networks, trained in an adversarial setting.
    A generator network, parameterized by a vector $\theta$, {$\mathcal{G}_\theta$}  produces some artificial data given some vector of random noise, and a discriminator network $\mathcal{D}$ differentiates between the synthetic output of the generator and the true data. Otherwise said, the generator network provides a way to map one distribution to another, and notably produces samples from the latter target distribution.
    
\medskip    
    The game between the generator $\mathcal{G}_\theta$ and discriminator $\mathcal{D}_\epsilon$ can be formally expressed as the minimax objective of distance $\mathcal{V}(\mathbb{P}_\mathrm{r},\mathbb{P}_\mathrm{g})$:
    \begin{align}
    	&\mathcal{V}(\mathbb{P}_\mathrm{r}, \mathbb{P}_\mathrm{g})  =\underset{\theta \in \mathbb{R}^{N_g}  }{\mathrm{min}} \; \underset{\epsilon \in \mathbb{R}^{N_d}} {\mathrm{max}} \; \nonumber\\ 
    	&\qquad\underset{\myvec{x} \sim \mathbb{P}_\mathrm{r}}{\mathbb{E}} \left[ \log \left( \mathcal{D}_\epsilon (\myvec{x}) \right) \right] + \underset{\myvec{z} \sim \mathbb{P}_\mathrm{z}}{\mathbb{E}} \left[ \log \left( 1 - \mathcal{D}_\epsilon(\mathcal{G}_\theta(\myvec{z})) \right) \right] ,
    	\label{eq:minimax_objective_GAN}
    \end{align}
	where $\mathbb{P}_{\mathrm{r}}$ corresponds to the data distribution and $\mathbb{P}_{\mathrm{g}}$ is the model distribution defined by the distribution $\mathbb{P}_{\mathrm{z}}$ transformed by the generator $\mathcal{G}_\theta$, with $\theta$ corresponding to the network weights.
	The source distribution $\mathbb{P}_{\mathrm{z}}$ is often a uniform or Gaussian distribution.
	In this work, the vector $\myvec{z}$, and the corresponding distribution $\mathbb{P}_{\mathrm{z}}$, are provided by another complex distribution, as discussed in Section~\ref{implementation_WGAN}.
	We note that the discriminator network must also be provided and optimized according to the weights $\epsilon$.

\medskip
    The training phase is completed when a Nash equilibrium \citep{nash1951} is reached, i.e. when neither of the two opponents can improve by unilaterally adjusting their strategy, and the discriminator cannot distinguish between the true and artificial data.
    At this point, the generator, in principle, would have learned to output a sufficiently good representation of the real data probability distribution, i.e. $\mathbb{P}_{\mathrm{g}} \approx \mathbb{P}_{\mathrm{r}}$, and would therefore be able to map a known or latent probability distribution to the target data distribution.
    
\medskip
	Unfortunately, the standard GAN framework is vulnerable to training instabilities, often resulting from issues involving vanishing gradients or mode collapsing, where the generator output lies in a restricted phase space, thereby producing incoherent results. 
    
    \subsection{Deep convolutional GANs} \label{DCGANs}
    
    Despite the drawbacks outlined above, GANs have still been shown to be extremely successful. \citet{radford2015unsupervised} developed an improved GAN architecture, known as deep convolutional GAN (DCGAN), by replacing the multilayer perceptrons in the generator and discriminator networks with convolutional layers \citep{lecun2015deep}.
    Other improvements, such as batch normalization \citep{ioffe2015batchnormalization}, i.e. ensuring that the input to each unit is normalized with zero mean and unit variance, were also introduced to stabilize the learning and promote gradient flow in deeper networks.
    Such infrastructural upgrades result in overall improved training stability and render the network more robust to discrete-mode and manifold model collapse \citep{metz2016unrolled,arjovsky2017towards}.
    Nevertheless, DCGANs remain susceptible to model instabilities.
    
    \subsection{Wasserstein GANs} \label{wasserstein_GANs}

\begin{figure*}
	\centering
		{\includegraphics[scale=0.625,clip=true]{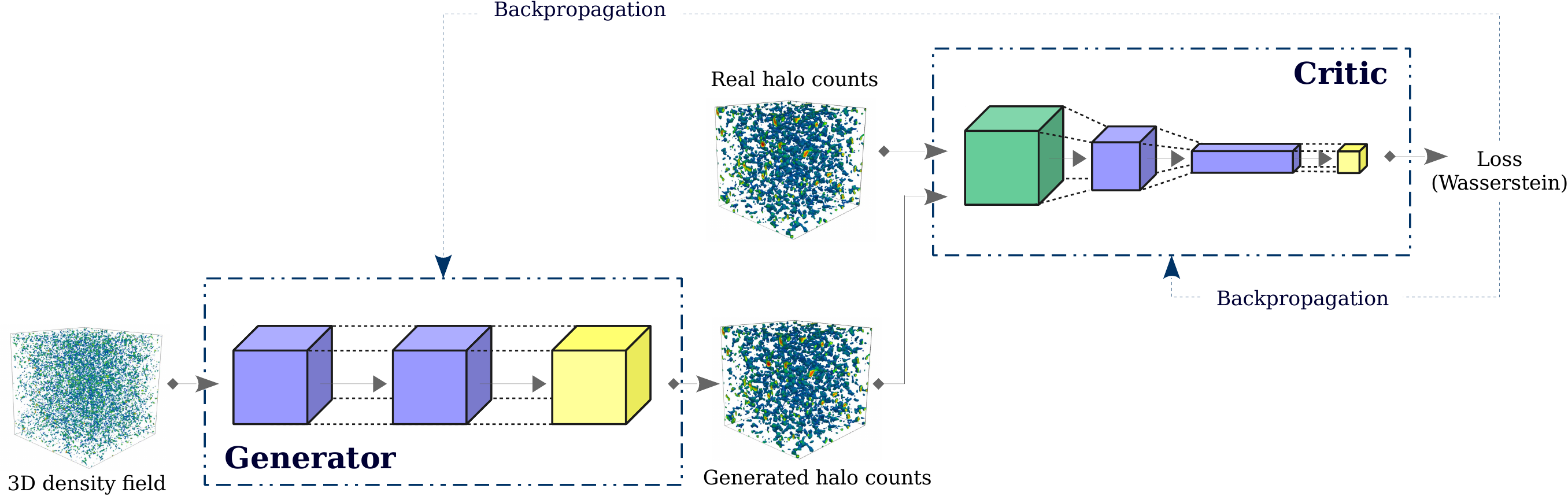}}
	\caption{Schematic representation of Wasserstein halo painting network implemented in this work. The role of the generator is to learn the underlying non-linear relationship between the input 3D density field and the corresponding halo count distribution. The difference between the output of the critic for the real and predicted halo distributions is the approximately learnt Wasserstein distance and is used as the loss function which must be minimized to train the generator.}
	\label{fig:WGN_schematic}
\end{figure*}

	\citet{arjovsky2017wgan} proposed another variant of GAN, which encodes an alternative loss function based on the Wasserstein-1 distance between a real and a generated distribution, also known as the Earth Mover's distance.
    This distance can be informally and intuitively interpreted as the minimum cost of transporting mass in order to transform a probability distribution into a given target distribution.
    This variant, referred to as Wasserstein GAN (or WGAN) despite there being no adversarial component to the training, has shown to be capable of learning arbitrarily complex probability distributions of a panoply of data sets \citep{arjovsky2017wgan}, leading to extremely realistic results \citep[e.g.][]{karras2017progressive}.
    
\medskip    
	The Wasserstein distance, $\mathcal{W} (\mathbb{P}_{\mathrm{r}}, \mathbb{P}_{\mathrm{g}})$, has the desired properties for the convergence of sequences of probability distributions \citep{arjovsky2017wgan}.
    The WGAN value function can be expressed via the Kantorovich-Rubinstein duality \citep{villani2008optimal} as
	\begin{equation}
	 \mathcal{W}(\mathbb{P}_\mathrm{r}, \mathbb{P}_\mathrm{g}) = \underset{\mathcal{C} \in \mathfrak{C}} {\mathrm{sup}} \; \left\{\underset{\myvec{x} \sim \mathbb{P}_{\mathrm{r}}}{\mathbb{E}} \left[ \mathcal{C}_\epsilon (\myvec{x}) \right] - \underset{\myvec{z} \sim \mathbb{P}_\mathrm{z}}{\mathbb{E}} \left[ \mathcal{C}_\epsilon(\mathcal{G}_\theta(\myvec{z})) \right] \right\},
	\label{eq:minimax_objective_WGAN}
	\end{equation}
	where the supremum is over the set of 1-Lipschitz functions denoted by $\mathfrak{C}$, such that minimizing the above value function with respect to the generator parameters will also minimize $\mathcal{W} (\mathbb{P}_{\mathrm{r}}, \mathbb{P}_{\mathrm{g}})$ for the case of an optimal discriminator.
    The discriminator network $\mathcal{D}(\myvec{x})$ is now designated as the critic $\mathcal{C}(\myvec{x})$, since it is not trained to differentiate or classify as in the standard GAN framework. 
    To ensure that the critic is a 1-Lipschitz function, the form of the function that the network takes must be closed.
    This condition can be enforced by restricting the allowed weight space for the critic network.
	WGANs are, therefore, simply generative networks where the loss function is learned using a second	network.
	This new distance mitigates the concept of adversarial training by broadening the concept further and allowing to measure differences between whole distributions. By relying on the earth-mover concept, it also ensures that the critic stays within a safe subspace contrary to GAN discriminatory network.
    A schematic representation of our generative network, used to establish a mapping between the 3D density field and its corresponding halo count distribution, is illustrated in Fig.~\ref{fig:WGN_schematic}.
    A detailed description of the training methodology is provided in Section~\ref{implementation_WGAN}.

    \subsection{Prior work involving GANs} \label{prior_work_GANs}
    
    GANs, and their variants mentioned above, are becoming increasingly popular among the astrophysical and cosmological community by virtue of their versatility and effectiveness to achieve impressive results.
    \citet{mustafa2017creating} developed a DCGAN to generate cosmological weak lensing convergence maps with high statistical confidence, while the de-noising of such maps within a GAN framework was investigated by \citet{shirasaki2018denoising}.
    GANs have also been optimized for the efficient generation of realistic 2D realizations of the cosmic web, demonstrating their ability to capture the complexity of large-scale structures \citep{rodriguez2018fast}.
    \citet{troster2019painting} employed a GAN to map dark matter density fields to gas pressure distributions, thereby augmenting $N$-body simulations with baryons.
    Other interesting applications of GANs involve de-noising galaxy images to recover impressive detailed features, outperforming standard convolution methods \citep{schawinski2017generative}, separating AGN from their host galaxy's light profile \citep{stark2018psfgan}, deblending galaxy superpositions \citep{reiman2018deblending}, atmospheric retrievals on exoplanets \citep{zingales2018exogan} and generating physically realistic galaxy images \citep{fussell2018forging} and deep galaxy fields \citep{smith2019generative}.
    The improved variant of WGAN has been used for the generation and refinement of signal patterns of particle detectors from simulations of cosmic-ray induced air showers \citep{erdmann2018generating}.
    \citet{zamudio2019HIgan} recently used a WGAN to generate 3D cosmic neutral hydrogen (HI) distributions with properties closely matching those from costly cosmological hydrodynamic simulations.

    \section{Network refinements} \label{network_refinements}

    To optimize the performance of our halo painting network, we consider several architectural upgrades which have been recently presented in the literature.
    In this section, we briefly describe the network refinements which are implemented in our network architecture and training machinery.
    
    \subsection{Inception} \label{inception}
    
    A particular deep convolutional architecture was proposed by \citet{szegedy2015going,szegedy2016rethinking}, code-named Inception, which achieved state-of-the-art performance for object classification and detection purposes.
    This is a novel level of organization which results in increased network depth and width, thereby improving the efficiency of deep neural network architecture.

\begin{figure}
	\centering
		{\includegraphics[scale=0.45,clip=true]{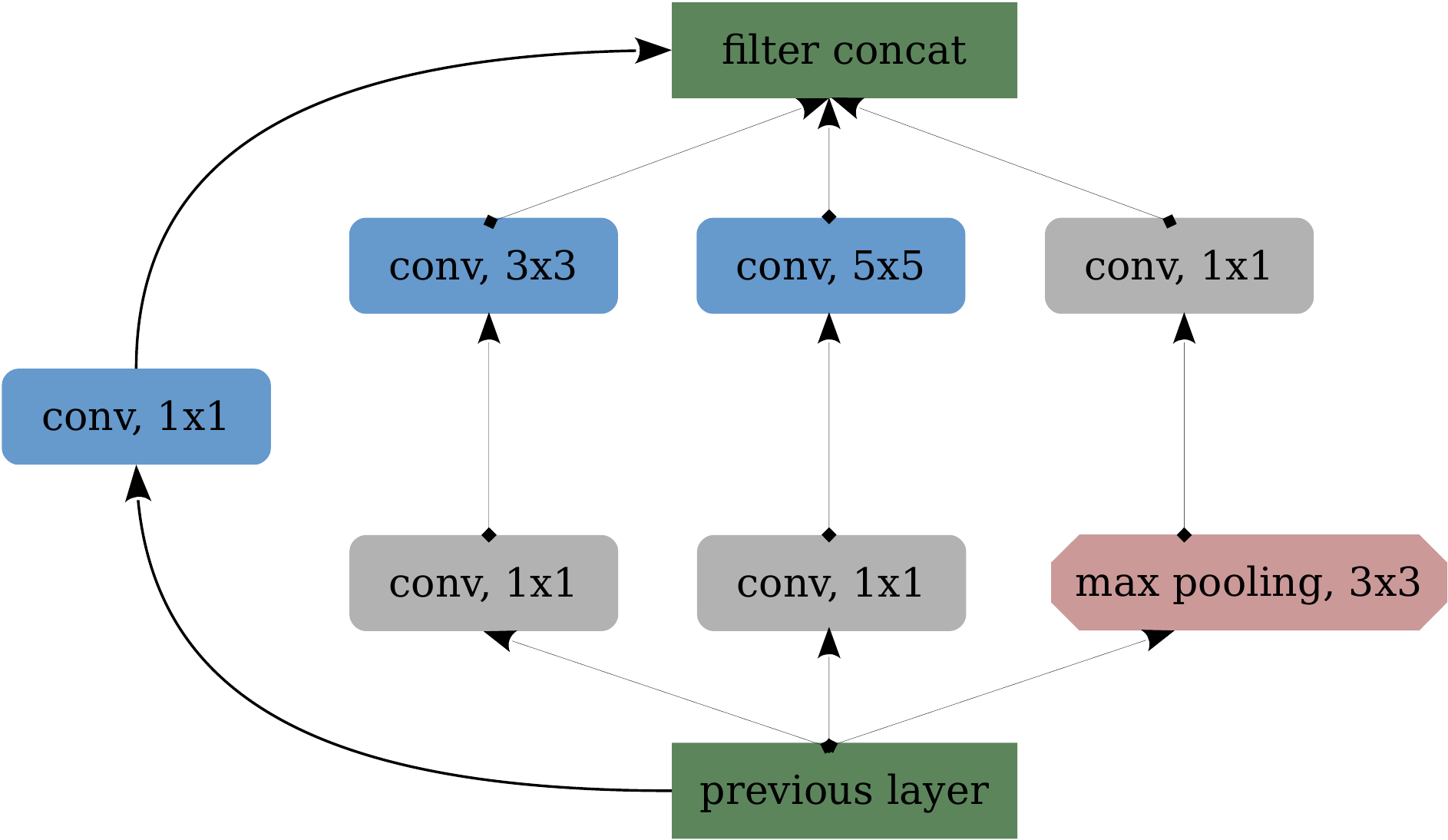}} 
	\caption{Original Inception block, with filter dimensionality reduction, as proposed by \citet{szegedy2015going}.
    The grey convolutional (``conv'') layers indicate the $1\times1$ convolutions introduced for the purpose of dimensionality reduction.
    We note that they are actually convolutions according along the feature space, of dimension $F$, of the previous layer, which is mono-dimensional, to produce $\tilde{F}$ new features but with the same physical size.
    As $\tilde{F} < F$, we achieve dimensionality reduction with respect to the number of features.
    The blue $1\times1$ convolutional layer corresponds to a normal convolution inside each feature.
    The output of the Inception module is the concatenation of the filters (``filter concat'') from the respective convolutional layers.
    This increases the dimension of the feature space depending on the number of output of each of the top convolutions.
    With the exclusion of the max pooling path, we make use of this architecture in the Wasserstein halo painting network.}
	\label{fig:inception_schematic}
\end{figure}

\medskip        
    The original Inception module \citep{szegedy2015going} consists of a series of convolutions, on the same level, with kernel sizes of $5\times5$, $3\times3$ and $1\times1$ in each Inception module, which is diagrammatically represented in Fig.~\ref{fig:inception_schematic}. 
    In this diagram, the blue box represents convolution within each feature space, while the gray convolutions are taken across features.
    In the case of blue convolutions, the kernel size defines the receptive field of a layer, i.e. the size of the patch of the input which affects the connected output.
    The outputs are concatenated and fed to the subsequent components of the network, which may be another Inception module.
    This concatenation increases the amount of features available for the next layer to make more computations.
    Stacking successive Inception modules yields further depth, with each module being optimized to recognize features on various scales.
    The Inception module, as proposed in \citet{szegedy2015going}, also introduces max pooling layers and $1\times1$ convolutions.
    A max pooling transformation downgrades the resolution of the input grid by some factor by taking local maxima.
    The two additional $1\times1$ convolutions are inserted in each branch before the $3\times3$ and $5\times5$ convolutions for dimensionality reduction to limit the computational resources required to a reasonable amount.
    Indeed, by convolving along the feature space, they may produce new compressed features with lower dimensions, though with the same size for each feature.
    This possibility of controlled dimensionality reduction is the crux of the Inception module. 

\medskip
    The essence of Inception lies in increased network depth, while obviating the potential drawbacks of deep convolutional networks.
    Na\"ively stacking large convolutional layers to build very deep networks is computationally expensive and renders the networks prone to over-fitting.
    Moreover, gradient updates may not flow smoothly throughout such networks.
    The key advantage of Inception is therefore a remarkable gain in quality where the computational workload is not greatly increased compared to networks with lower depth and width.
    
    \subsection{Residual connections} \label{resnet}

\begin{figure}
	\centering
		{\includegraphics[scale=0.45,clip=true]{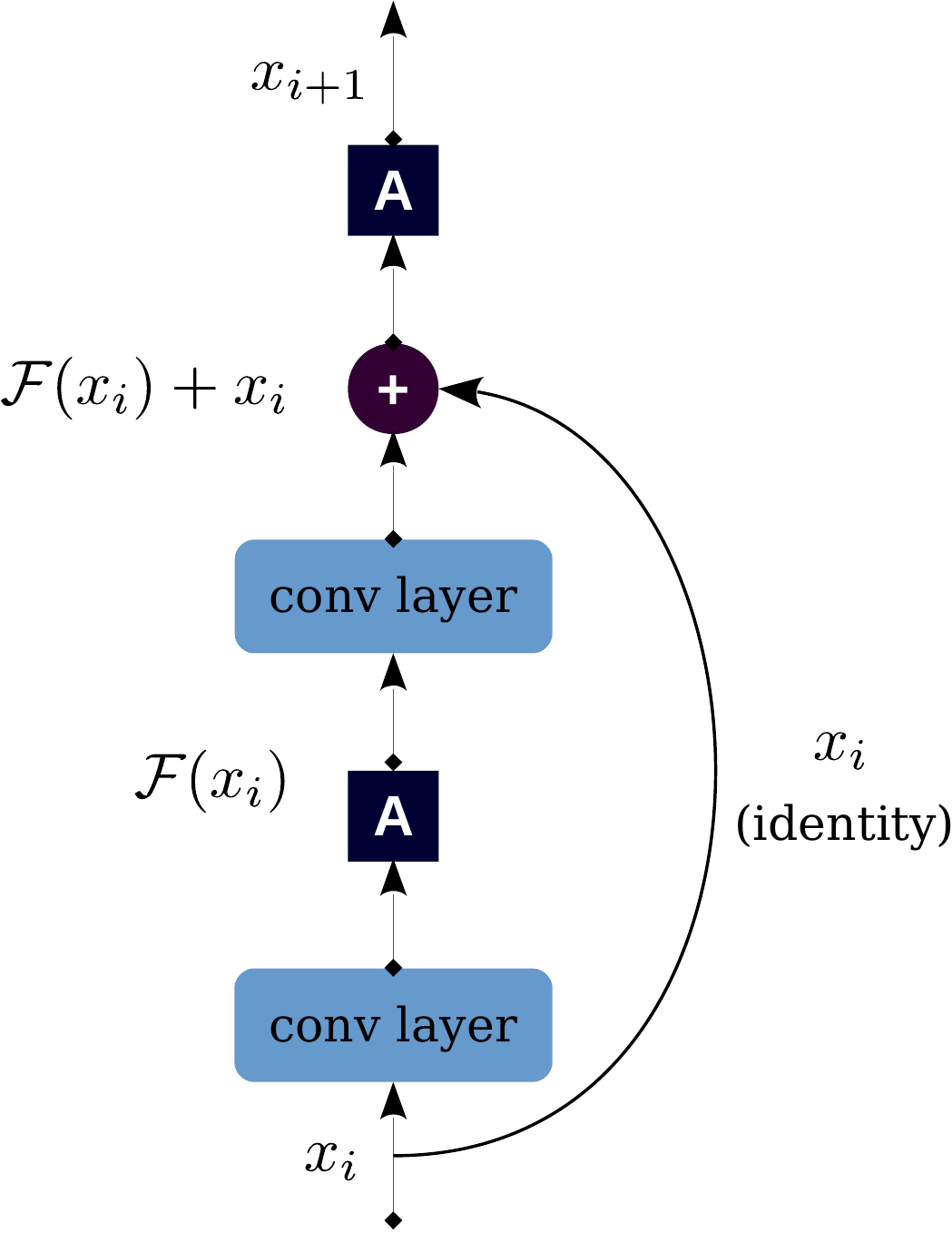}} 
	\caption{Residual learning block, as proposed by \citet{he2016resnet}.
    Activations are denoted by {\bf A}, with element-wise addition indicated by the plus ($\bm{+}$) symbol.
    In this case, the residual learning is implemented via an identity shortcut connection which skips two convolutional layers.
    As such, the output of the above ResNet is $x_{i+1} = \mathcal{F}(x_i) + x_i$, where $\mathcal{F}(x_i)$ represents the residual mapping.
    In the case where $x_i$ and $\mathcal{F}(x_i)$ do not possess the same dimensions, an additional convolutional layer without any activation function is introduced before the sum.}
	\label{fig:resnet_schematic}
\end{figure}

    A residual learning framework was proposed by \citet{he2016resnet} to improve the training of deep neural networks.
    Conceptually, the framework relies on a reformulation of a given layer as learning the residual mapping with reference to its input, rather than directly learning the desired underlying mapping.
    \citet{he2016resnet} empirically demonstrated that such residual networks mitigate the degradation issue, whereby very deep networks are susceptible to saturation and eventually degradation of the training accuracy.
    This is the result of higher training error with increasing network depth. Residual networks are commonly abbreviated as {\it ResNets}.
    
\medskip        
    The desired underlying mapping can be formally denoted as $\mathcal{H}(x_i)$, such that the stack of non-linear layers would fit the residual mapping of $\mathcal{F}(x_i) = \mathcal{H}(x_i) - x_i$, where $i$ labels a given layer.
    The original mapping can therefore be reformulated as $\mathcal{H}(x_i) = \mathcal{F}(x_i) + x_i$.
    The hypothesis is that the residual function of $\mathcal{F}(x_i)$ is easier to optimize than the desired function of $\mathcal{H}(x_i)$ \citep{he2016resnet}.
    The formulation of $\mathcal{F}(x_i) + x_i$ can be implemented via feedforward neural networks with ``shortcut connections''.
    Such connections skip one or more layers and perform identity mapping.
    
\medskip        
    The implementation of residual connections in typical deep neural networks is straightforward.
    The only modification involves adding an extra identity shortcut connection between the input to a given layer and the output of the subsequent layer (cf. Fig.~\ref{fig:resnet_schematic}).
    Another positive aspect of adding residual connections is that no extra parameters are required and therefore the computational workload is not greatly increased except for negligible element-wise addition.
    The residual learning framework, in a nutshell, yields substantial accuracy gains by allowing extremely deep networks to be trained since the gradient information can flow without degradation to the early layers, thereby alleviating the vanishing gradient problem.
    In a standard encoder-decoder structure, this especially facilitates the propagation of small-scale information as the size of the images is reduced gradually in the encoding phase \citep{isola2016image}.
    
\medskip
    In this work, we implement a combination of ResNets and Inception architecture, as depicted in Fig.~\ref{fig:residual_inception_schematic} with the same color convention as in Fig.~\ref{fig:inception_schematic}, yielding residual Inception blocks.
    \citet{szegedy2017residualinception} have demonstrated that the introduction of residual connections within the Inception module leads to significant improvement in training speed.
    The state-of-the-art performance obtained by combining these two network refinements is a crucial factor behind our choice of network architecture for the halo painting network.
    
    \subsection{Gradient penalty} \label{gradient_penalty}
    
    As discussed above, the weights of the critic network for WGANs must be restricted to ensure that this network is 1-Lipschitz.
    When first conceived, weight clipping was used to enforce this criterion.
    However, \citet{gulrajani2017improved} demonstrated that the use of weight clipping can lead to poor performance of WGANs in certain scenarios, such as optimization difficulties, which may be mitigated via batch normalization \citep{ioffe2015batchnormalization}, although this does not guarantee convergence of very deep WGAN critics as illustrated in their work.
    \citet{gulrajani2017improved} therefore came up with an alternative in the form of a gradient penalty in the loss function.
    This obviates the undesirable behaviour induced by weight clipping, while yielding substantial performance improvements.
    
\medskip
	Since a differentiable function is 1-Lipschitz if and only if the norm of its gradient is at most 1 everywhere, \citet{gulrajani2017improved} proposed to directly constrain the gradient norm of the critic's output with respect to its input.
    The Lipschitz constraint is hence imposed by penalizing the gradient norm for random samples $\hat{\myvec{x}} \sim \mathbb{P}_{\hat{\myvec{x}}}$, where $\hat{\myvec{x}} = \epsilon \myvec{x} + (1 - \epsilon) \tilde{\myvec{x}}$ and $\epsilon$ is sampled randomly and uniformly, $\epsilon \in [0,1]$, resulting in the following augmented objective:
	\begin{multline}
	 \mathcal{L} =  \underset{\myvec{z} \sim \mathbb{P}_\mathrm{z}}{\mathbb{E}} \left[ \mathcal{C}_\epsilon(\mathcal{G}_\theta(\myvec{z})) \right] - \underset{\myvec{x} \sim \mathbb{P}_{\mathrm{r}}}{\mathbb{E}} \left[ \mathcal{C}_\epsilon (\myvec{x}) \right]  \\ + \lambda \underset{\hat{\myvec{x}} \sim \mathbb{P}_{\hat{\myvec{x}}}}{\mathbb{E}} \left[ \left( || \nabla_{\hat{\myvec{x}}} \mathcal{C}_\epsilon (\hat{\myvec{x}}) ||_2 - 1 \right)^2 \right] ,
	\label{eq:augmented_loss_gradient_penalty_GAN}
	\end{multline}
	where $\lambda$ is an arbitrary penalty coefficient and $\lambda = 10$ has been shown to work well for a range of architectures and data sets \citep{gulrajani2017improved}.
    Essentially, we must introduce a gradient penalty term in the original critic loss, which forces the gradient of the critic network to remain close to unity.
    Alternativately, interpreting the loss function as the opposite of the log-likelihood, this term states that not more than a fluctuation of $\sim \sqrt{1/10}$ compared to one is allowed for the norm of the gradient.
    A key advantage of the gradient penalty alternative is that it yields stable gradients which allows training of deep and complex networks without requiring {\it ad hoc} solutions such as batch normalization.

    \section{\textsc{velmass} simulation} \label{velmass}

	In this work, we adopt the reference element of the \textsc{velmass} cosmological simulation suite.
    The \textsc{velmass} suite is comprised of 10 cosmological simulations, 9 of which are probing slightly different variations of a selection of cosmological parameters whilst using the same initial phases.
    The $10^\textrm{th}$ simulation has the same parameter values as the central simulation described below, but with different initial phases, such that it can be independent from the other simulations, allowing us to perform blind model comparison.
    The simulation that we use in this work assumes a Planck-like cosmology \citep{13planck2015} with $\Omega_\text{m}=0.315$, $\Omega_\text{b}=0.049$, $H_0=68\,\mathrm{km~s}^{-1}\mathrm{ Mpc}^{-1}$, $\sigma_8=0.81$, $n_{\mathrm{s}}=0.97$ and $Y_\mathrm{He}=0.248$ (named ``central'' or $\Omega$ simulation).
    The power spectrum is obtained through the analytic prescription of \citet{eisenstein1999power}, and the initial conditions were generated by \textsc{music} \citep{hahn2011music}. 
    This simulation suite is designed to test the robustness of analysis tools to acceptable variations in cosmology.
    
    \medskip
    The cosmological simulation covers a volume of 2000$h^{-1}$~Mpc with 2048$^3$ particles tracing dark matter.
    It was initialized at a redshift $z=50$ and evolved to present time with \textsc{gadget2} \citep{springel2005gadget2}, adopting a softening length for gravity equal to 48$h^{-1}$~kpc corresponding to $1/20$ of the mean interparticle separation.
    The \textsc{rockstar} halo finder algorithm \citep{behroozi2013rockstar} was subsequently employed to extract the halos from the simulation and generate the 3D halo field. The halo identification involves finding particles 
    belonging to regions for which the local density is above a specific threshold, in our case as derived by the Friend-Of-Friend linking length, 45 times the mean density. Sub-halos are found by reducing recursively the linking length to find more compact structure. Each halo/sub-halo is pruned if it holds less than 10 particles as they are considered to be unstable. \textsc{rockstar} further performs a test for each particle to check if it is gravitationally bound to the structure. If this is not the case, the particles are removed from the halo. Once all these procedures are done, we are left with 23\% of the total mass in structures considered as virialized by \textsc{rockstar}.
    We histogram the halo counts onto a grid of 512$^3$ which is the resolution we choose to work at, i.e. each voxel has a side length of $L \approx 4h^{-1}$~Mpc.
    The halos were selected in four equally spaced logarithmic bins in the mass range: $10^{12} - 10^{14}$ $h^{-1}$~$\mathrm{M}_{\odot}$.
    Finally, we also produce the result of a pure second order Lagrangian perturbation theory (2LPT) simulation by setting the redshift for which \textsc{music} must create initial conditions to $z=0$.
    We build the corresponding gridded density fluctuation field by applying the cloud-in-cell (CIC) algorithm \citep[e.g.][]{hockney1988computer} to the particle distribution at the same resolution that we chose for the halo distribution, i.e. each voxel with length $L\approx 4h^{-1}$~Mpc.
    
    \medskip
    We can test the cosmological dependence of the halo painting network by complementing the $\Omega$ simulation with other variants obtained assuming different cosmological parameters, for testing purpose. 
    The initial random phases are kept the same and only the effects induced by different cosmological parameters is introduced. 
    In this work, we concentrate on two additional $2048^3$ cosmological simulations with  $\Omega_\textrm{m}=0.355$ and $\Omega_\textrm{m}=0.275$. 
    We repeat the entire procedure, above, for these two simulations, i.e. execution of the $N$-body simulation, identification of halos, construction of the 2LPT field and both resampling the halos and performing the CIC of the 2LPT field onto a 512$^3$ grid.
    
    \section{Model architecture and training methodology} \label{implementation_WGAN} 

\begin{figure}
	\centering
		{\includegraphics[scale=0.45,clip=true]{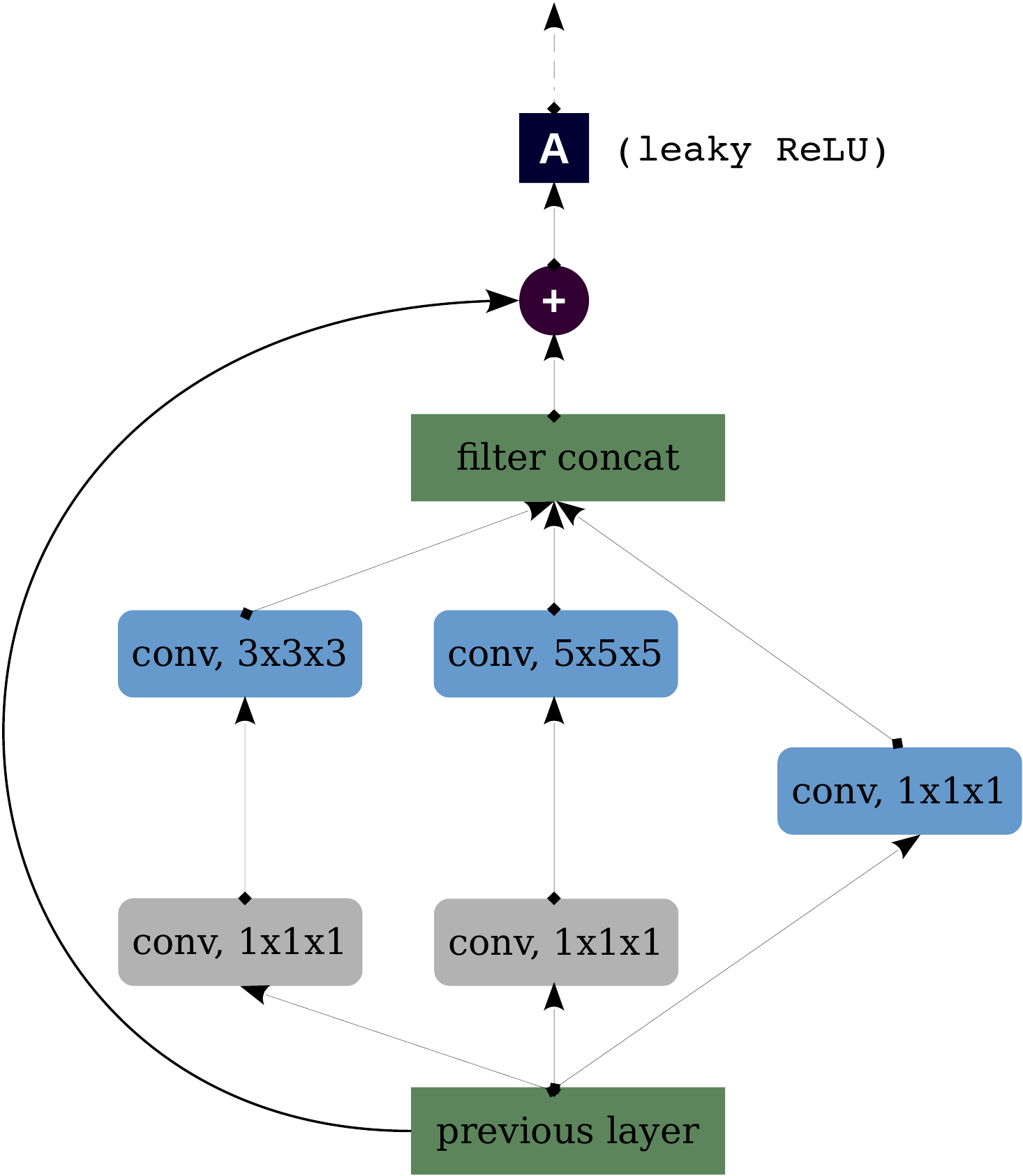}} 
	\caption{Residual Inception block, as implemented in this work, for the halo painting network architecture.
    Our Inception module is a slightly modified version of the original architecture (cf. Fig.~\ref{fig:inception_schematic}).
    In particular, we replace all 2D convolutional layers with their 3D counterpart and the only dimensionality reduction comes from not padding the input when performing the convolutions.
    All convolutional layers employ 10 filters, with a {\ttfamily leaky ReLU} activation implemented after the residual connection.}
	\label{fig:residual_inception_schematic}
\end{figure}

\begin{figure*}
	\centering
		{\includegraphics[scale=0.625,clip=true]{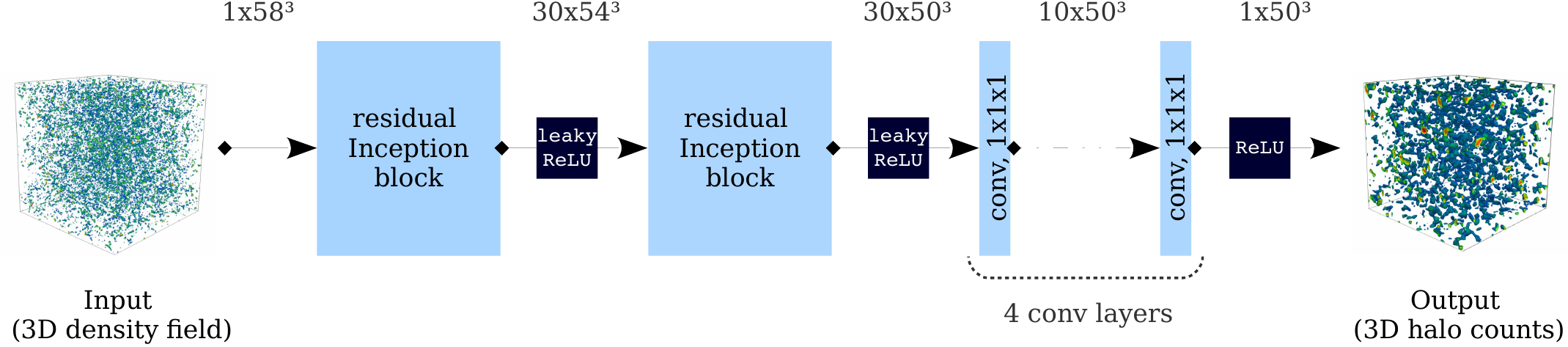}}
	\caption{Schematic representation of the halo painting network implemented in this work.
    The network consists of two residual Inception blocks (cf. Fig.~\ref{fig:residual_inception_schematic}), followed by a series of four convolutional layers with $1\times1\times1$ kernel, across feature space as in the gray box in the Inception module.
    The activation function used is a leaky rectified linear ({\ttfamily leaky ReLU}) unit, except for the output layer where a rectified linear ({\ttfamily ReLU}) activation ensures non-negative halo counts in the generated 3D halo field.
    The input to the halo mapping network is a 3D realization of density field, with the output being the corresponding halo count distribution.
    The input is conveniently chosen to be larger to eliminate the need for padding.
    In the schematic, we indicate the size of the tensors used during training. We can use any input density fields sampled on a mesh with side $N_\textrm{density}>25$ which will predict a halo field sampled on a smaller mesh with side $N_\textrm{halo}=N_\textrm{density}-8$.}
	\label{fig:generator_schematic}
\end{figure*}

\begin{figure*}
	\centering
		{\includegraphics[scale=0.625,clip=true]{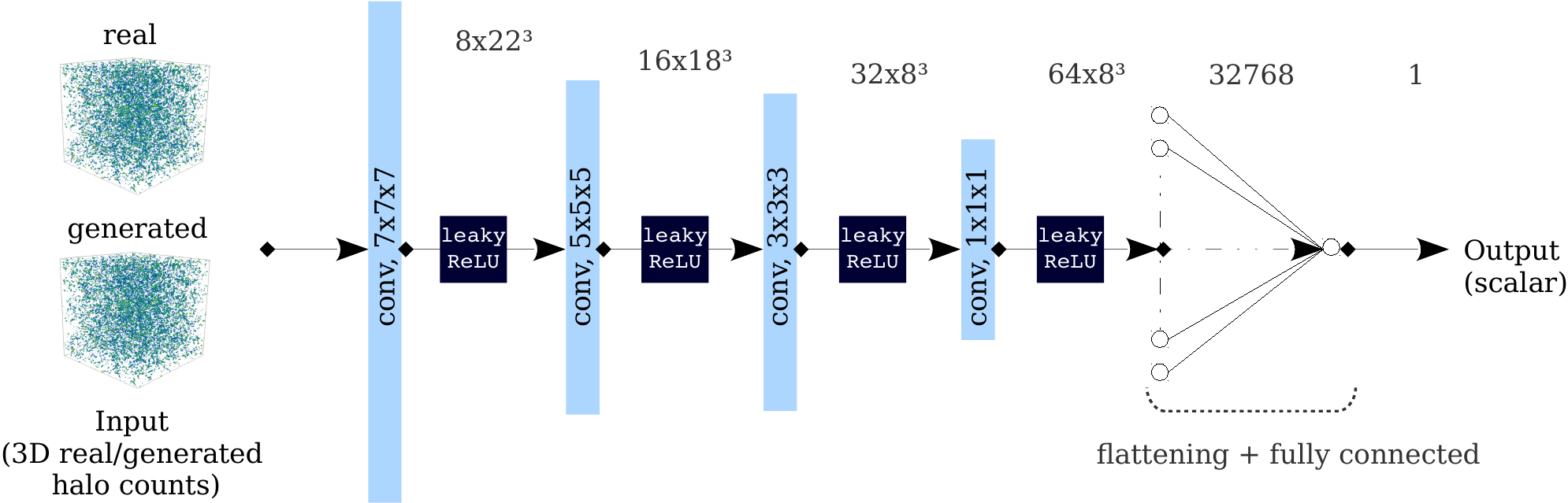}}
	\caption{Schematic representation of the critic employed in this study.
    The network encodes a series of four convolutional layers, gradually reducing their respective kernel sizes from $7\times7\times7$ to $1\times1\times1$, and activated with {\ttfamily leaky ReLU}, with the corresponding output flattened and fed into a fully connected layer with linear activation.
    The critic reduces the input real and generated halo fields each to a compact scalar representation whose difference is an approximation to the Wasserstein distance.}
	\label{fig:critic_schematic}
\end{figure*}

\begin{figure}
	\centering
		{\includegraphics[width=\hsize,clip=true]{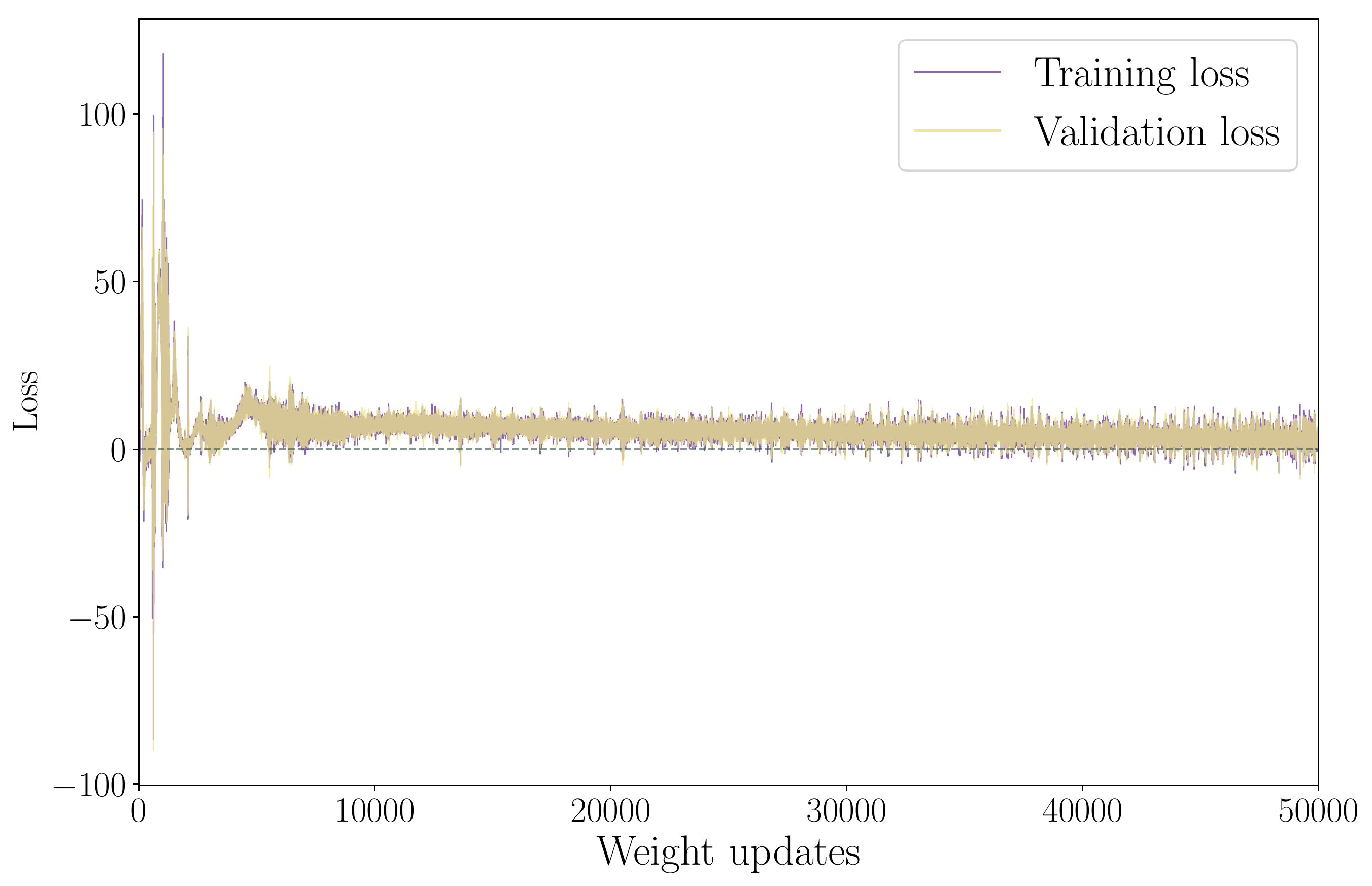}}
	\caption{Training and validation loss for our physical mapping network for the first $5\times10^4$ weight updates.
    These respective losses are an approximation to the Wasserstein distance, as measured by the critic.
    As expected, this distance tends to zero as training proceeds and our halo painting network effectively learns the mapping from the dark matter to halo distribution. We note that the validation loss is closely tracking the training loss. As it can be seen, the two curves are mostly overlapping in the above plot, except at the very edge.}
	\label{fig:training_validation_loss}
\end{figure}

    Our halo painting network is built to perform the map between the dark matter distribution and the halo count distribution.
    This allows us to use physical intuition to guide the model architecture.
    Considering the halo mapping network, we input a patch of the 2LPT field directly, unlike the usual conception of a GAN where a flat array of noise is normally used as the input.
    We are therefore physically mapping from a well-understood distribution of dark matter instead of a latent space.
    Although our halo painting network is not a generator in the classical machine learning sense, we refer to it as the ``generator'' below to make the analogy with the WGAN training routine. The ``generator'' terminology is also justified because the mapping network transforms a complex density field sampled from a Gaussian random field transformed by the 2LPT dynamics into a halo field with even less trivial statistical properties.
    
\medskip    
    Knowing that information from the dark matter field is aggregated from a relatively local patch via some non-linear process allows us to choose a simple form for the generator network, a schematic of which is shown in Fig.~\ref{fig:generator_schematic}.
    We connect the local region of the dark matter field using a 3D modification of the residual Inception block, shown in Fig.~\ref{fig:residual_inception_schematic}.
    Since the receptive field of $5\times5\time5$ convolution kernel is $~20h^{-1}$Mpc per side, then performing a second $5\times5\times5$ convolution on the output of the first increases the receptive field to $\sim 40h^{-1}$Mpc per side.
    In fact, since the kernels are 3D, the furthest distance that information can be propagated from the dark matter field to the halo distribution is $70h^{-1}$Mpc across the diagonal of such stacked kernels, whilst still learning the small scale features from the $1\times1\times1$ kernels and the mid-scale features with the $3\times3\times3$ kernels.
    This distance is far enough that the halo field should be insensitive to influences at such a scale.
    Since we use residual connections in the Inception blocks, we combine structure from distant patches, whilst still retaining a close relation to the density field itself.
    To learn the non-linear process from this receptive density field, we use four convolution layers with kernel size $1\times1\times1$ with no residual connection, which provides the non-linearity necessary to combine the local density distribution and perform the map to the distribution of halo counts.
    Since there is an enormous complexity in the 2LPT field, we use many filters in each layer to learn the wide variety of possible features. 
    Every single kernel in our network has 10 filter channels to provide an extremely large path to build the complex non-linear map.
    For such a generator, we have 31,930 trainable parameters which is relatively few (in machine learning terms).
    The non-linearity is provided by the popular {\ttfamily leaky ReLU} activation function with a leaky parameter of $\alpha=0.1$.
    To ensure the positivity of the halo count field, we use {\ttfamily ReLU} at the last layer.
    
\medskip
    In principle, we can train the generator network using 2LPT density patches as small as 9$^3$ voxels to predict a single halo count distribution voxel (due to using convolutions with no padding), although the learned kernels, and therefore the predicted halo count field, would not be sensitive to any information from outside of this region.
    The maximum extent of the generator comes from the size of the kernels in the two residual Inception blocks, which is 25$^3$ voxels of the density field to predict a 17$^3$ halo patch.
    To increase the number of features to attempt to learn at once, we actually select 58$^3$ voxels from the gridded 2LPT field to predict a 50$^3$ voxel halo count distribution.

\medskip
    A major advantage to this prescription of building the generator in such a way is that we can train the network using small simulations and predict massive halo fields provided with large, cheap 2LPT fields.
    For example, we could, in parallel, make millions of 58$^3$ voxel cosmological simulations and run the halo finder on each for the training data, which is relatively quick in comparison to running an extremely large, say 2048$^3$ voxel simulation.
    Then, using the trained generator, we could predict \emph{any} size halo field just by providing the 2LPT calculation, which is relatively cheap compared to performing the same sized simulation.
    Such a large 2LPT density field slab with $508\times508\times58$ voxels is used to predict a $500\times500\times50$ halo distribution slab and the projection is shown in Fig.~\ref{fig:projected_halo_predictions} for the central slice of depth $\sim 100h^{-1}$~Mpc and side length of $\sim2000h^{-1}$~Mpc.
    
\medskip
    Provided with our generator network which during training, as described above, will take 58$^3$ voxel 2LPT fields to paint 50$^3$ voxel halo count distributions, we now need to build a critic network to measure the distance between the painted halo counts and the corresponding real halo counts from the \textsc{velmass} simulation.
    Our critic, as depicted in Fig.~\ref{fig:critic_schematic}, utilizes a series of four convolutional layers, while gradually reducing their respective kernel sizes from $7\times7\times7$ to $1\times1\times1$, and activated with {\ttfamily leaky ReLU} ($\alpha = 0.1$), with the output of the last convolutional layer flattened and fed into a fully connected layer with linear activation.
    The critic encodes relevant information from the input real or predicted halo fields into compact representations, thereby reducing the size of the 3D distributions.
    The output of the critic is a single scalar which is used to compute the approximately learned Wasserstein distance between the predicted and true halo distributions given a particular generative network.
    This output can therefore be used to compute the loss function {\eqref{eq:minimax_objective_WGAN}} which is minimized to train the generative network.
    We implement the above networks and the training routine outlined below in TensorFlow \citep{abadi2016tensorflow}.

\medskip
    During training, we load both the entire 2LPT density field and histogrammed halo count distribution from the $\Omega$ \textsc{velmass} simulation into the TensorFlow graph and select by index sub-volume elements of size 58$^3$ and 50$^3$, respectively, which massively reduces computation time compared with passing the 3D slices of data at each weight update.
    These patches corresponds to side lengths of $L\approx225h^{-1}$~Mpc and $L\approx200h^{-1}$~Mpc, respectively.
    The input to the generator is randomly chosen and the corresponding true halo counts volume is selected. 
    Note that here, the batch size is unity, such that the number of weight updates corresponds to the number of density field patches used to train the generator.
    We use a $512^3$ simulation box for training, where we use a large portion of the box for training and the remaining section for validation, such that we utilize non-mutual parts of the box for the training and validation set.
    To encode some further symmetries through our training set, we also perform a rotation of the selected patches, thereby extracting the input 3D slice from a randomly oriented region.
	The generator employs the gradient of the Wasserstein loss function (\ref{eq:minimax_objective_WGAN}) with respect to its parameters $\theta$ for training.
    
\medskip
	The initial training step involves the optimization of the weights of the critic network to minimize the augmented loss function (\ref{eq:augmented_loss_gradient_penalty_GAN}), while concurrently freezing the parameters of the generator.
	The weights of the critic must be updated $n_{\mathrm{critic}}$ times, where $n_\mathrm{critic}$ is sufficient for the critic to converge.
	The samples for this initial step are randomly selected (and rotated) 2LPT fields and corresponding true halo counts.
    In the subsequent step, the critic weights are temporarily anchored, and the generator parameters are adjusted.
    The training routine then proceeds in iterative fashion, until an overall convergence of the generator is achieved.
    The training rationale is to reduce the Wasserstein distance between the true halo counts and the halo counts mapped from the corresponding input density field such that the generator gradually learns the correct mapping.
    The training procedure outlined above is represented schematically in Fig.~\ref{fig:WGN_schematic}.
    
\medskip
    In this work, we use $n_{\mathrm{critic}} = 5$ and set the arbitrary coefficient for the gradient penalty to $\lambda = 10$, with $\epsilon$ having a randomly and uniformly drawn value $0 \leq \epsilon \leq 1$.
    We implement the popular {\it Adam} \citep{kingma2014adam} optimization algorithm, with a learning rate of $10^{-4}$ and first and second moment exponential decay rates of 0.5 and 0.999, respectively.
    We trained the network for $\sim 5\times10^5$ generator weight updates for the different mass bins, requiring around 30 hours on a NVIDIA Quadro P6000.
    The training and validation loss for the first $5\times10^4$ weight updates of our halo painting model is illustrated in Fig.~\ref{fig:training_validation_loss}.
    This is the approximately learnt Wasserstein distance which tends to zero as training proceeds and the generator effectively learns the mapping from the dark matter to halo distribution.

    \section{Results} \label{results_WGAN}
    
\begin{figure*}
	\centering
		{\includegraphics[width=\hsize,clip=true]{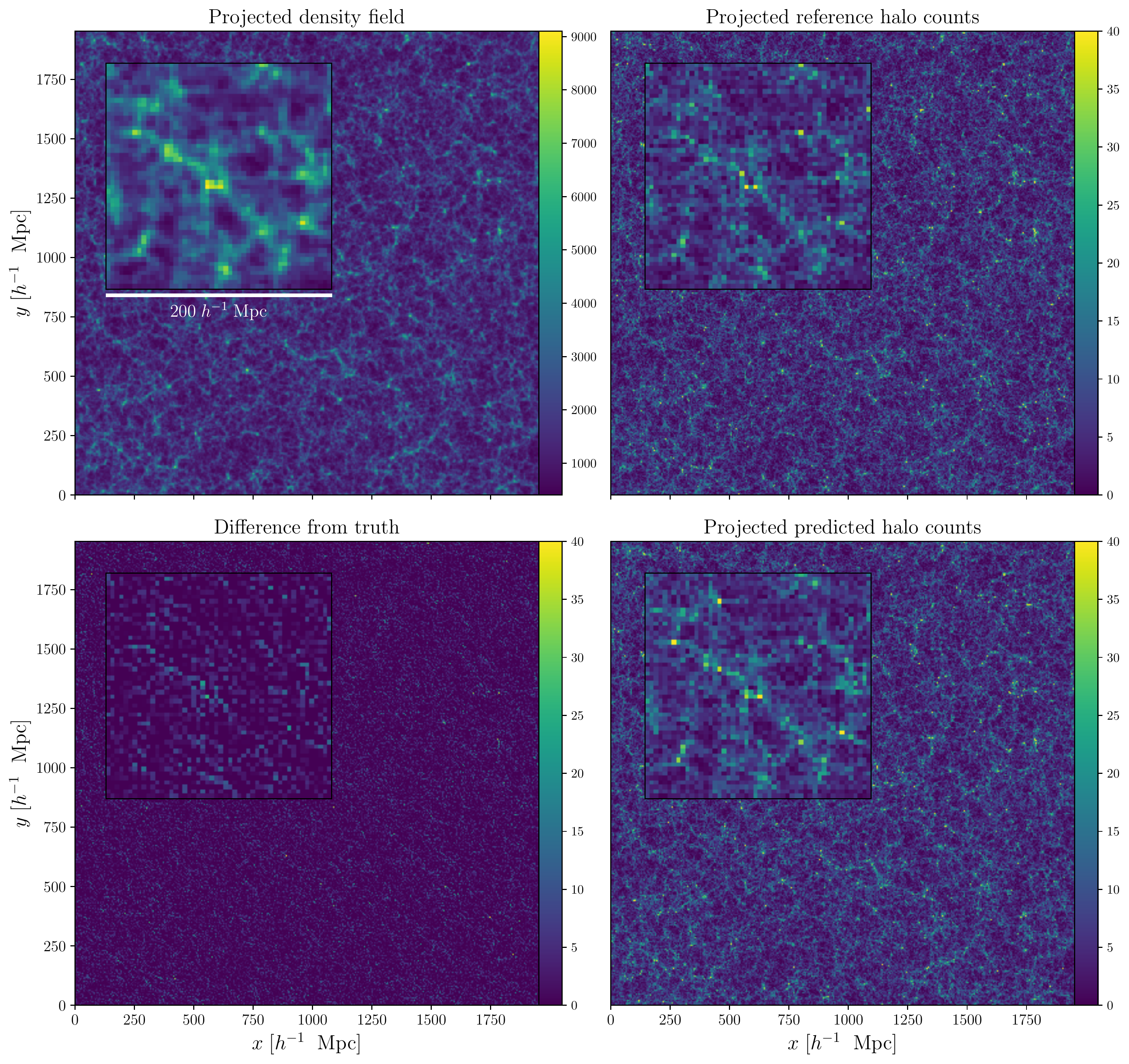}} 
	\caption{Prediction of 3D halo field by our halo painting model for a slice of depth $\sim 100h^{-1}$~Mpc and side length of $\sim2000h^{-1}$~Mpc.
    A blind validation dataset is shown in the top right panel, with the predicted halo count depicted below it.
    The corresponding 2LPT density field is displayed in the top left panel, with the difference between the reference and predicted halo distributions depicted in the lower left panel.
    A visual comparison of the reference and predicted halo count distributions indicates qualitatively the efficacy of our halo painting network.
    Note that we did not normalize the 2LPT density field in this work, which renders the performance of our network even more remarkable.}
	\label{fig:projected_halo_predictions}
\end{figure*}    

\begin{figure*}
	\centering
		{\includegraphics[width=\hsize,clip=true] {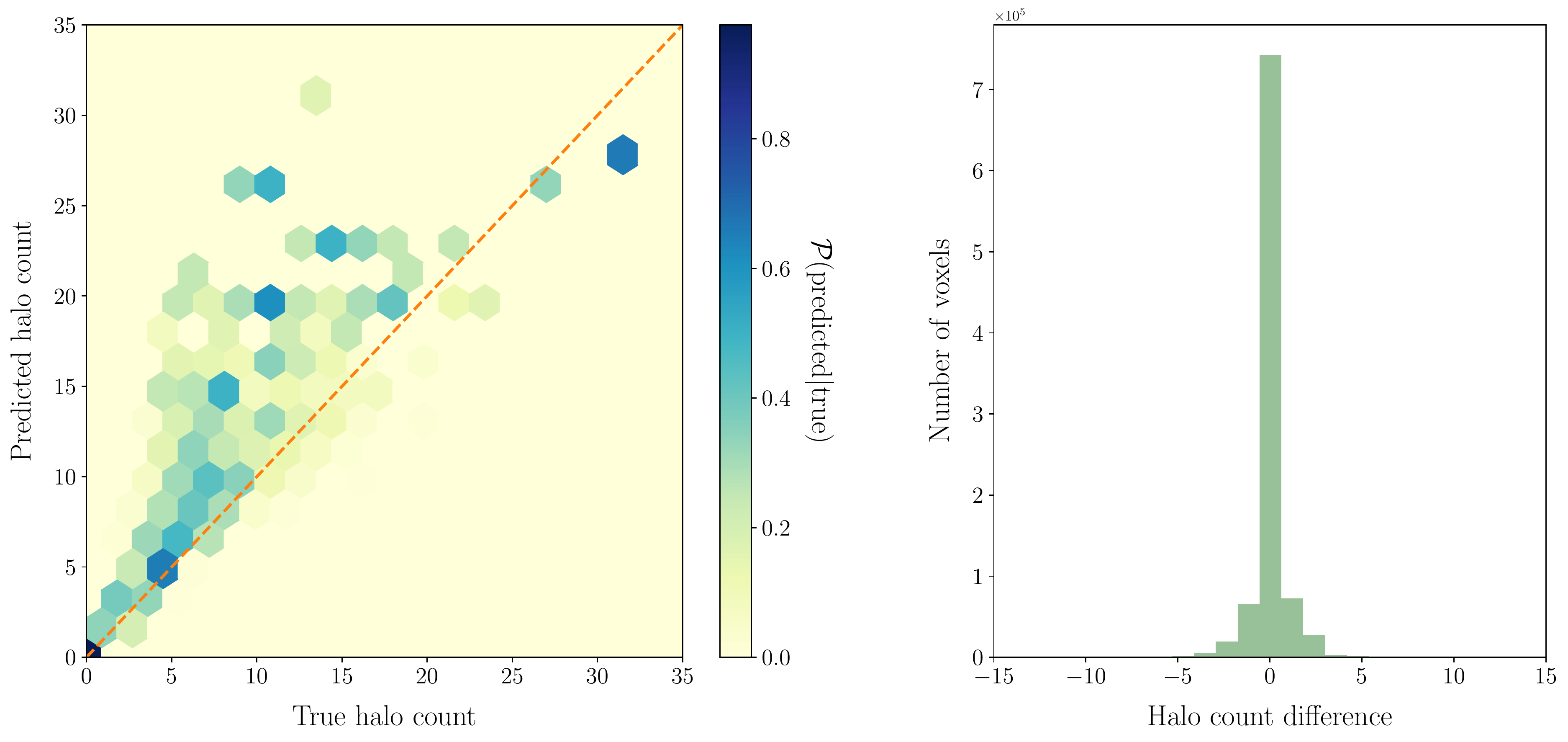}}
	\caption{{\it Left panel:} The conditional probability distribution of the predicted halo count per voxel given the corresponding true (i.e. reference) value. This properly represents the error in our prediction, while accounting for the intrinsic halo distribution.
	Our network predictions fare well in lower and average density environment with halo counts less than 10, but overshoots when the halo count is higher.
    There are, however, very few regions which reach such high halo count ($\lesssim 0.03\%$), such that our predictions are not often skewed by such rare occurrences.
    {\it Right panel:} The distribution of the difference between the reference and predicted halo counts per voxel. This shows that the difference is close to zero for the majority of the voxels, with the difference being larger than 3 for only $\sim 1.5\%$ of the total number of voxels.}
	\label{fig:conditional_probability_hist}
\end{figure*}

\begin{figure*}
	\centering
		{\includegraphics[scale=0.5,clip=true] {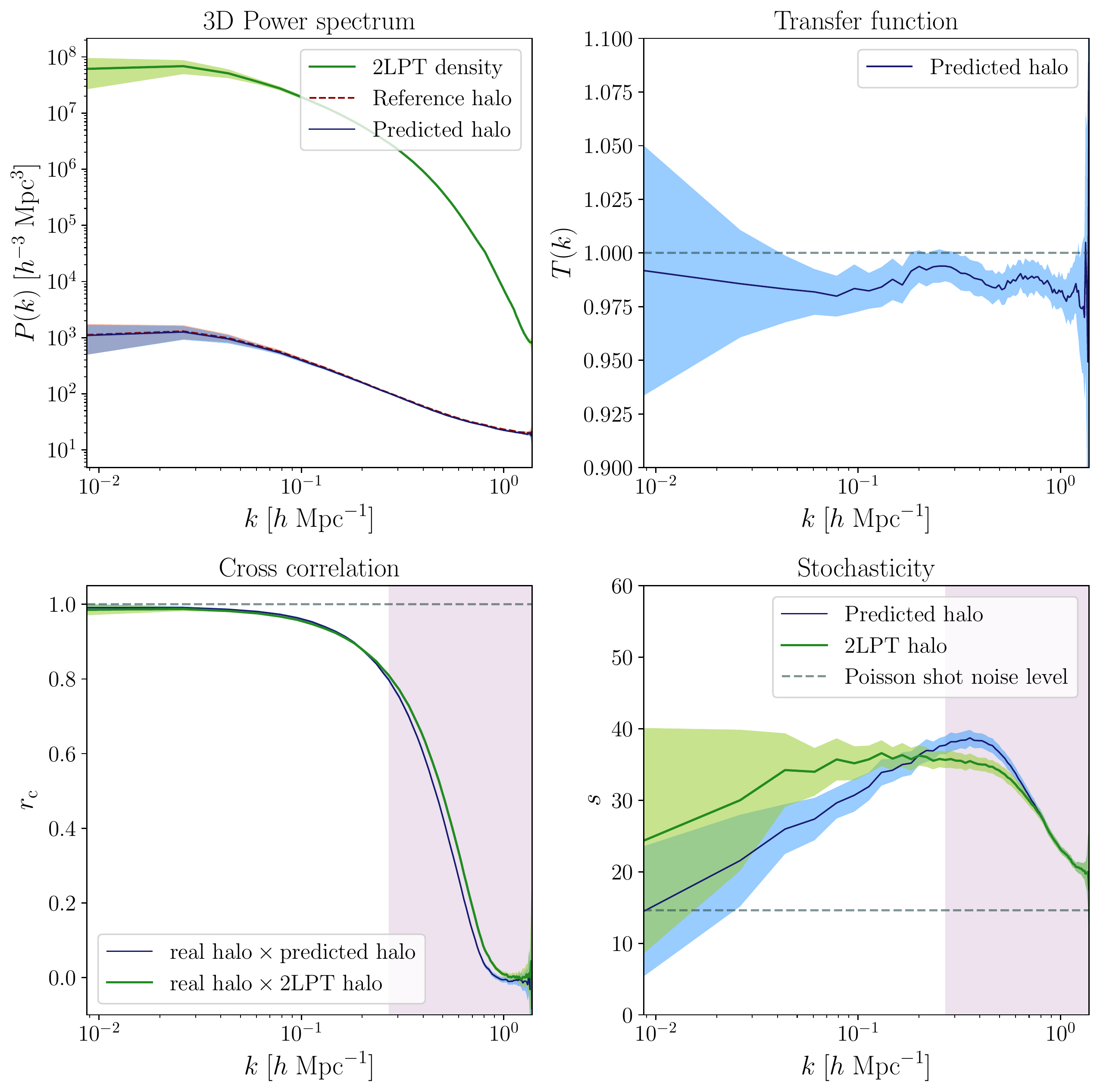}} 
	\caption{The summary statistics of the 3D power spectra of the density, reference and predicted halo fields for one thousand randomly selected patches.
    The solid lines indicate their respective means, while the shaded regions indicate their respective $1\sigma$ confidence regions, i.e. 68\% probability volume.
    The above diagnostics demonstrate the ability of our halo painting model to reproduce the characteristic statistics of the reference halo fields and therefore provide substantial quantitative evidence for the performance of our neural network in mapping 3D density fields to their corresponding halo distributions.
    Note that the turnover in the cross-correlation coefficient is due to the grid resolution, and the prediction in the shaded area is redundant in practice.
    The behaviour of the cross-correlation coefficient and stochasticity of the halo field derived linearly from 2LPT, illustrated in the bottom panels, shows that our network predictions reproduces the reference halo field with overall higher fidelity, at least up to the regime limited by the grid resolution.}
	\label{fig:summary_statistics_combined_bins}
\end{figure*}

\begin{figure*}
	\centering
		{\includegraphics[width=\hsize,clip=true] {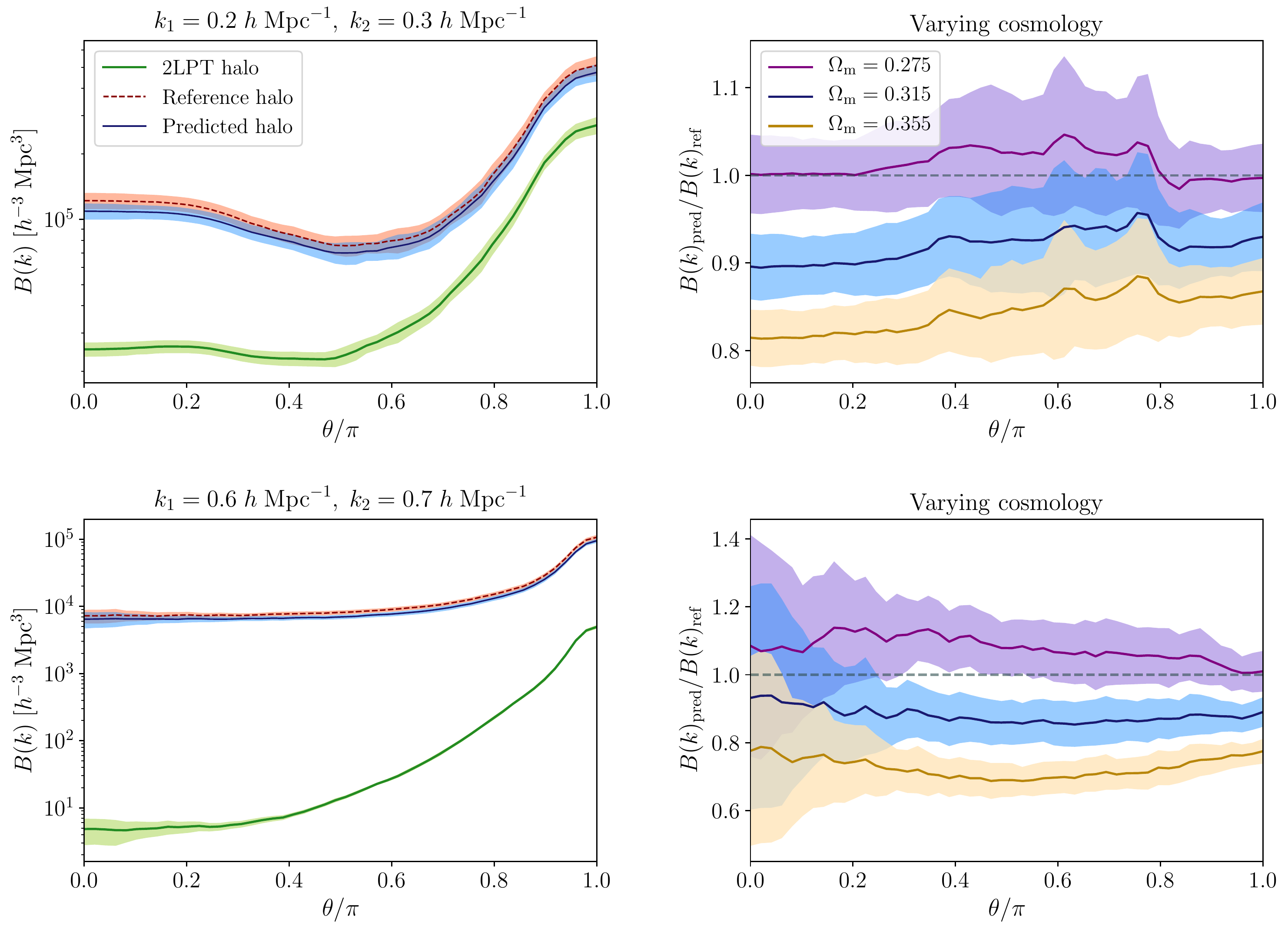}}
	\caption{{\it Left panels:} The summary statistics of the 3D bispectra of the 2LPT, reference and predicted halo fields for a given small- and large-scale configurations, $k_1 = 0.6 \; h \; {\mathrm{Mpc}}^{-1}$ and $k_2 = 0.7 \; h \; {\mathrm{Mpc}}^{-1}$, $k_1 = 0.2 \; h \; {\mathrm{Mpc}}^{-1}$ and $k_2 = 0.3 \; h \; {\mathrm{Mpc}}^{-1}$, respectively.
	In both cases, there is a close agreement between the bispectra from the reference and predicted halo distributions. 
	In particular, our network predictions are a significant improvement over the corresponding 2LPT halo fields.
	{\it Right panels:} The deviation from the 3D bispectra of the reference halo distributions of the corresponding predictions for the two cosmology variants.
	The above bispectrum diagnostics show that our network is more sensitive to the fiducial cosmology than at the level of the two-point correlation function.
	The $1\sigma$ confidence regions for five hundred randomly selected patches are depicted in each panel.}
	\label{fig:bispectrum_cosmo_variation}
\end{figure*}

\begin{figure*}
	\centering
		{\includegraphics[width=\hsize,clip=true] {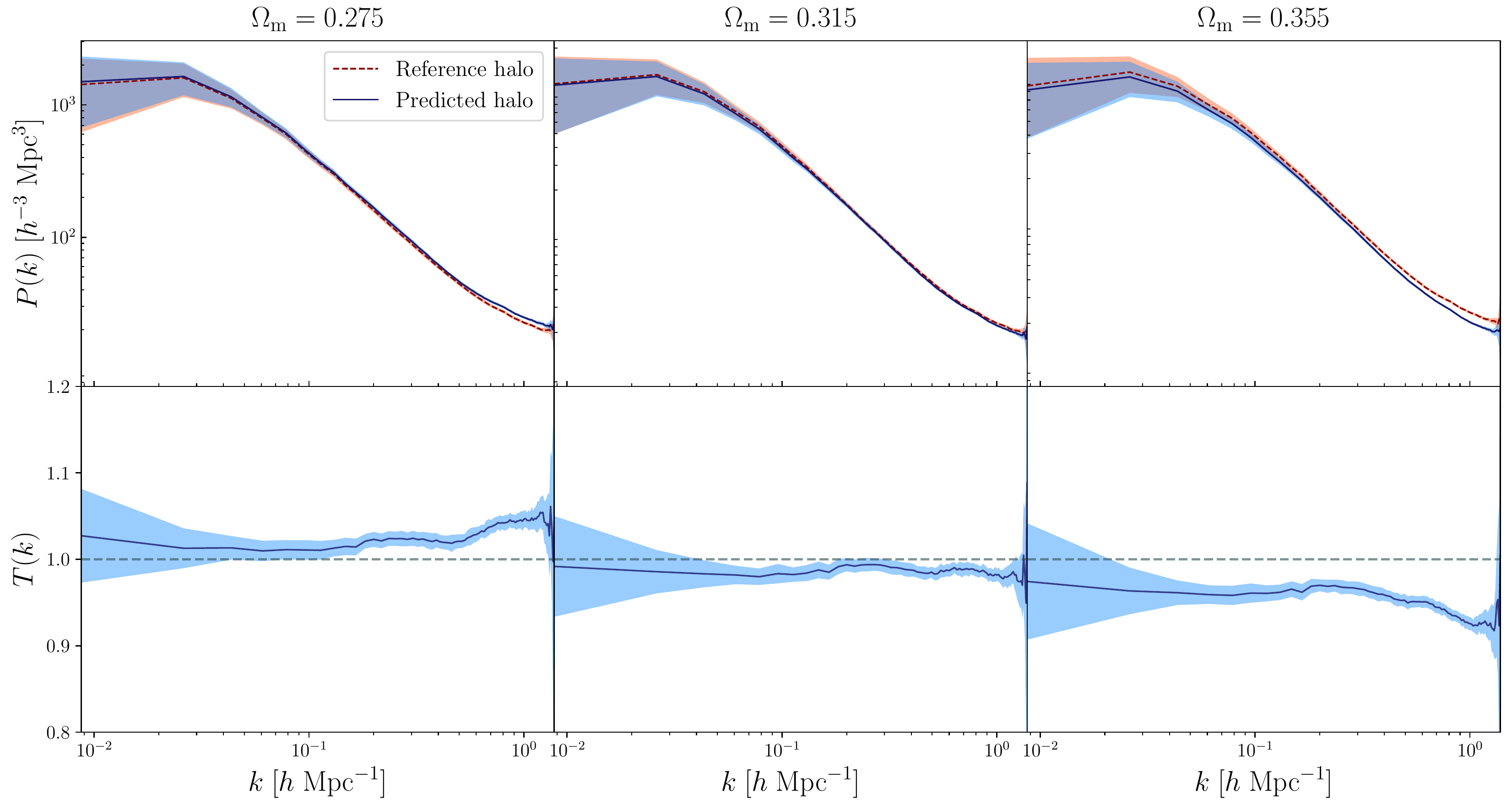}}
	\caption{{\it Top panels:} The corresponding power spectra, as in the top left panel of Fig.~\ref{fig:summary_statistics_combined_bins}
	, for the fiducial and two cosmology variants, with $1\sigma$ confidence regions for one thousand randomly selected patches.
	The top left and right panels demonstrate the capability of our halo painting network to reproduce the two-point summary statistics when applied to simulations generated with the fiducial cosmology.
    {\it Bottom panels:} The corresponding transfer functions highlight the consistency between the power spectra reconstructed from the predicted and real halo fields for the three cosmology variants, with the deviation from their respective reference spectra being below 10\%.
    The above diagnostic therefore shows that our halo painting model is slightly sensitive to the underlying cosmology at the level of the power spectrum.}
	\label{fig:summary_statistics_combined_cosmo}
\end{figure*}

    Fig.~\ref{fig:projected_halo_predictions} depicts the reference and predicted halo fields, for a 3D slice of depth $\sim 100h^{-1}$~Mpc and side length of $\sim2000h^{-1}$~Mpc, and the difference between the reference and predicted fields, obtained from a completely unseen simulation.
    The corresponding 2LPT density field is also shown for the sake of completeness.
    Qualitative agreement is impressive, implying that the halo painting network is capable of mapping the complex structures of the cosmic web, such as halos, filaments and voids, to the corresponding distribution of halo counts.
    
    \medskip
    We illustrate the conditional probability distribution of the predicted halo count per voxel given the respective true (or reference) value in the left panel of Fig.~\ref{fig:conditional_probability_hist}, with the right panel depicting the distribution of the difference between the reference and predicted values.
    The conditional probability distribution, which accounts for the intrinsic halo distribution, indicates that our predictions fare well in lower and average density environment (counts less than 10) while it overshoots when the halo count is high (counts greater than 10).
    But it is important to note that there are very few regions which reach such high halo count ($\lesssim 0.03\%$) and thus our network predictions are not often skewed by such rare occurrences.
    The distribution of the difference shows that our network predictions for most voxels closely match the corresponding true values, with the difference being larger than 3 for only $\sim 1.5\%$ of the total number of voxels.
    We now assess and validate the performance of our halo painting model using other quantitative diagnostics.
    
    \subsection{Two-point correlation and power spectrum} \label{two_point_correlation_WGAN}

    As quantitative assessment, we employ summary statistics, as per the standard practice in cosmology.
    These summary statistics provide a reliable metric to gauge our halo painting network in terms of their capacity to encode essential information.
    
\medskip
    The two-point correlation function, denoted by $\xi(r)$, is the quintessential measure employed by cosmologists, along with its Fourier transform, the power spectrum $P(k)$, defined as follows:
\begin{align}
	\xi(|\myvec{r}|) &= \langle \delta(\myvec{r}') \delta(\myvec{r}' + \myvec{r}) \rangle \label{eq:2PCF_WGAN} \\
	P(|\myvec{k}|) &= \int \mathrm{d}^3 \myvec{r} \; \xi(\myvec{r}) e^{i\myvec{k}\cdot\myvec{r}} , \label{eq:power_spectrum_WGAN}
\end{align}
    where the $\delta$'s correspond to the field relative contrast, i.e. $\delta(\myvec{r}) = \rho(\myvec{r})/\bar{\rho}-1$, with $\rho(\myvec{r})$ the matter density and $\bar{\rho}$ the mean matter density.
    Assuming the cosmological density field is approximately a Gaussian random field, as is the case on the large scales or at earlier times, the above two statistics provide a sufficient description of the field.
    
\medskip    
    We first consider the power spectra of the reference and predicted halo distributions, denoted by $P_{\rm{hh}}(k)$ and $P_{\rm{pp}}(k)$, as a standard measure for the description of the matter and halo distributions.
    The indices $\rm{p}$ and $\rm{h}$ label the predicted and true (reference) halo modes, respectively.
    The top left panel of Fig.~\ref{fig:summary_statistics_combined_bins} illustrates their respective mean and $1\sigma$ confidence regions for one thousand randomly selected patches with side lengths of $L\approx400h^{-1}$~Mpc from a blind set of data, thereby quantitatively showcasing the remarkable performance of the halo painting model in learning the map from 2LPT density fields to halo number counts.
    
\medskip
	We also investigate some standard diagnostics at the level of two-point correlation functions in Fourier space to evaluate the model performance against the ground truth.
    The three metrics considered, each dependent on the mode $k$, are as follows:
    \begin{align}
        \intertext{1. Transfer function ($T(k)$)}
        T(k) &\equiv \sqrt[]{\frac{P_{\rm{pp}}(k)}{P_{\rm{hh}(k)}}}
	\label{eq:transfer_function_GAN}\\
    	\intertext{2. Cross correlation coefficient ($r_{\mathrm{c}}$)}
    	r_{\mathrm{c}}(k) &= \frac{P_{\rm{hp}}(k)}{\sqrt[]{P_{\rm{hh}(k)} P_{\rm{pp}}(k)}} 
	\label{eq:cross_correlation_coeff_GAN}\\
    	\intertext{3. Stochasticity ($s(k)$)}
        s(k) &= P_{\rm{hh}}(k) ( 1 - r_{\mathrm{c}}(k)^2) , \label{eq:stochasticity_GAN}
    \end{align}
    where, as before, $P_{\rm{hh}}(k)$ and $P_{\rm{pp}}(k)$ are the auto power spectra of the reference and predicted halo fields, respectively, while $P_{\rm{hp}}(k)$ corresponds to the cross power spectrum of the reference and predicted halo field. 
	The transfer function, defined as the square root of the ratio of the two auto power spectra, indicates the discrepancy in amplitudes, as a function of the Fourier modes, whilst the cross correlation coefficient characterizes the mismatch in phases between the real and predicted halo count fields.
	These two diagnostics quantify the predictive capability of our halo painting network, while the fraction of the variance that cannot be accounted for in the true model is encoded in the stochasticity.

\medskip
	Fig.~\ref{fig:summary_statistics_combined_bins} displays the scale dependence of the above summary statistics for the reference and predicted halo count fields with one thousand independent randomly selected patches, from an unseen simulation, in terms of their respective mean and 1$\sigma$ confidence regions.
    The power spectra of the predicted halo fields match extremely closely that of the reference halo distributions.
    The cross correlation coefficient is close to unity on the large scales and drops to $r_c\approx0.7$ at $k\approx0.4$.
    The turnover, at $k\approx0.25$, is due to the grid resolution, and the prediction in the shaded area is not taken into account in practice.
    The above two-point summary statistics showcase the  performance of our network in predicting the entire halo distribution.
    We also make a comparison with the corresponding diagnostics of the 2LPT halo field which corresponds to a statistical description of the halo distribution, derived from the 2LPT density field, which is valid, by construction, at the two-point level and on large scales.
    It is built by scaling the 2LPT density field by the adequate linear bias factor so that we match the power spectrum of the halo field on large scales.
    The behaviour of the cross-correlation coefficient and stochasticity of the 2LPT halo field, illustrated in the bottom panels of Fig.~\ref{fig:summary_statistics_combined_bins}, shows that our network predictions reproduces the reference halo field with overall higher fidelity, at least up to the regime limited by the grid resolution.
    On large scales, the ensemble mean stochasticity is lower by typically a factor of two, and its variance is smaller by about 30\%.
    On small scales, which are not covered by our predictor, the two estimates behave mostly in a similar way.
    Over the entire scale range, our predictor fails to go below the level given by the mean Poisson shot noise.
    We will investigate in the future how to improve the current situation, particularly on large scales as shot noise should not be a limitation given that the entire procedure is fully deterministic, both on the $N$-body simulation and the neural network sides.

    \subsection{Three-point correlation and bispectrum} \label{three_point_correlation_WGAN}

\begin{figure*}
	\centering
		{\includegraphics[scale=0.5,clip=true] {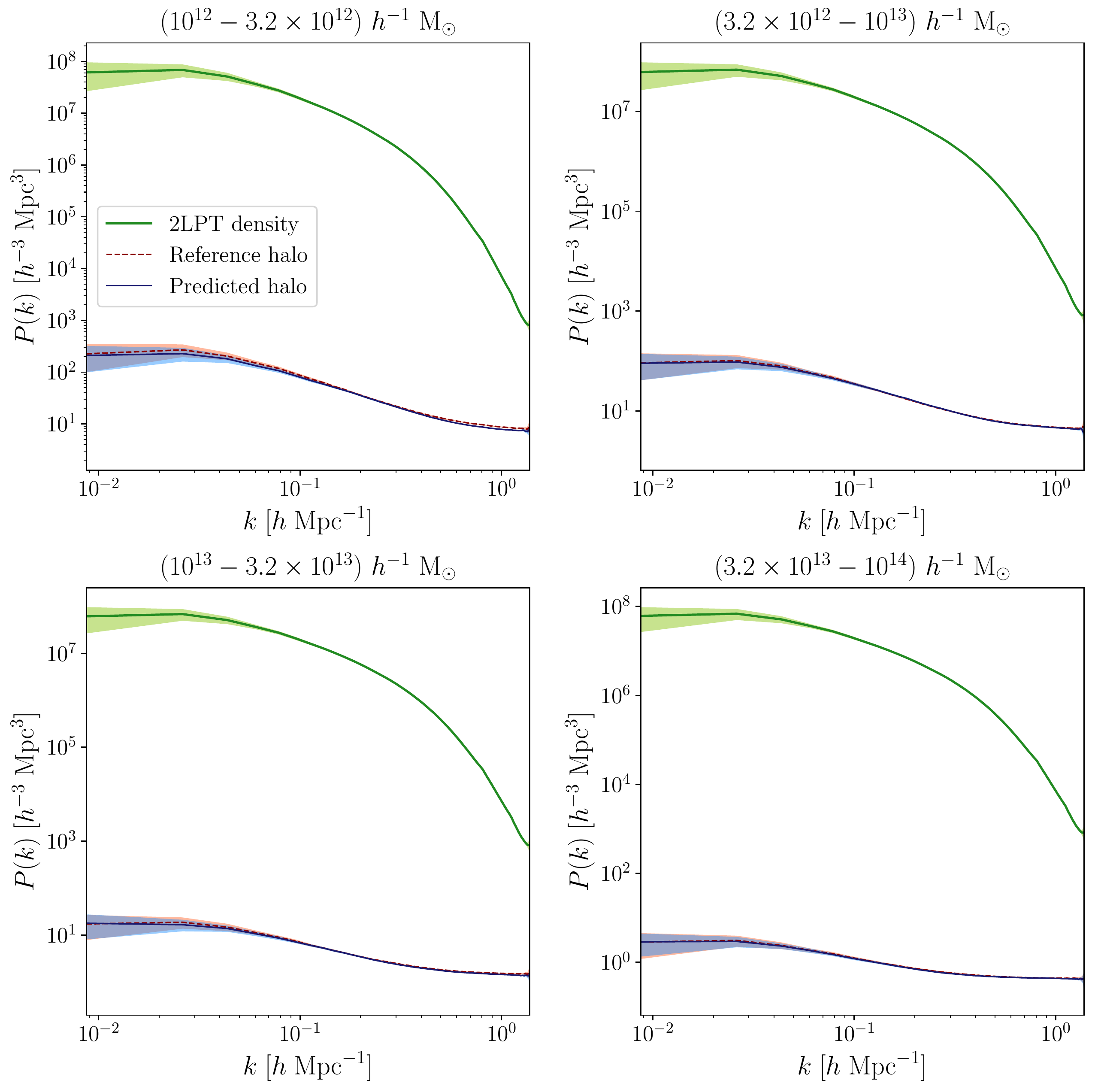}}
	\caption{The corresponding power spectra, as in the top left panel Fig.~\ref{fig:summary_statistics_combined_bins}, for the individual mass bins, with $1\sigma$ confidence regions for one thousand randomly selected patches.
	The above diagnostic, at the level of two-point statistics, highlights the remarkable performance of our halo painting network across the range of halo masses involved.
	As such, it can also predict the mass distribution of halos.}
	\label{fig:summary_statistics_individual_bins}
\end{figure*}

\begin{figure*}
	\centering
		{\includegraphics[width=\hsize,clip=true] {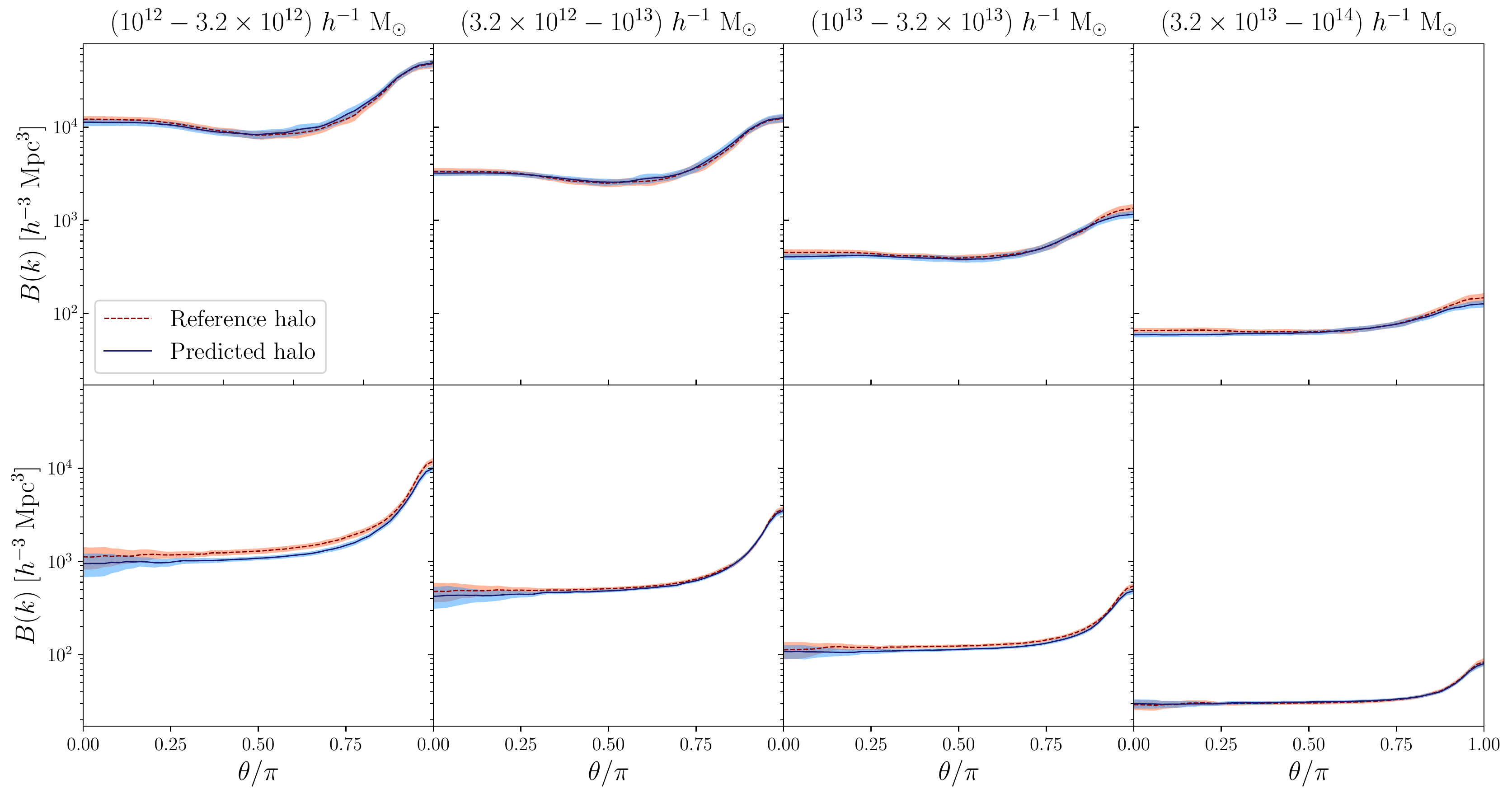}}
	\caption{The bispectrum diagnostics for the individual mass bins, for five hundred randomly selected patches.
	The top and bottom panels depict the large and small configurations, respectively, as considered in the left panels of Fig.~\ref{fig:bispectrum_cosmo_variation}.
	At the level of three-point statistics, we find that our network performs remarkably well across the range of halo masses involved.}
	\label{fig:bispectrum_individual_bins}
\end{figure*}

    The non-linear dynamics involved in gravitational evolution of cosmic structures contributes to a certain degree of non-Gaussianity of the cosmic density field on the small scales. Higher-order statistics are therefore required to characterize this non-Gaussian field. 
    We employ the bispectrum, i.e. the Fourier transform of the three-point correlation function, to quantify the spatial distribution of the density field, defined as: 
    \begin{equation}
    	(2\pi)^3 B(\myvec{k}_{1}, \myvec{k}_{2}, \myvec{k}_{3}) \delta_{\mathrm{D}}(\myvec{k}_{1}+\myvec{k}_{2}+\myvec{k}_{3}) = \langle \delta(\myvec{k}_{1}) \delta(\myvec{k}_{2}) \delta(\myvec{k}_{3}) \rangle , \label{eq:bispectrum_WGAN}
    \end{equation}    
    where $\delta_{\mathrm{D}}$ is the Dirac delta.
    The bispectra reconstructed from the 2LPT, reference and predicted halo fields are displayed in Fig.~\ref{fig:bispectrum_cosmo_variation}.
    In particular, we show the bispectra for a given small- and large-scale configurations.
    The construction of the 2LPT halo field is described above in Section \ref{two_point_correlation_WGAN}.
    This allows us to make a fair comparison between the clustering of the respective halo fields.
    The left panel demonstrates that our halo painting network reproduces the non-linear halo field both on the small and large scales, and is therefore capable of mapping the complex cosmic structures apparent in the reference halo field.
    The agreement between the reference and predicted halo bispectra on the largest scales is slightly suboptimal as both our halo painting and critic networks are not tailored to produce structure on this level.
    Our network predictions also show a significant improvement over the corresponding 2LPT halo fields.
    The above bispectrum computations were performed using the publicly available \textsc{pylians}\footnote{Available from \url{https://github.com/franciscovillaescusa/Pylians}} code.

    \subsection{Dependence on cosmology} \label{robustness_cosmo_WGAN}

    We investigate the influence of the fiducial cosmology adopted for the simulations on the efficacy of our halo mapping model.
    We depict in Fig.~\ref{fig:summary_statistics_combined_cosmo} the network predictions for two cosmology variants in terms of their respective two-point summary statistics.
    The corresponding transfer functions, depicted in the bottom panels, show a deviation of about 10\% from the reference power spectra of their respective real halo distributions on the smallest and largest scales.
    The right panels of Fig.~\ref{fig:bispectrum_cosmo_variation} illustrate the ratio of the predicted to reference bispectra for the two configurations displayed in the left panels.
    We find that there is a more significant dependence of our network on the fiducial cosmology at higher order statistics.
    
\medskip    
    The above diagnostics therefore demonstrate a certain degree of sensitivity on the underlying cosmology, as expected, since the non-linearly evolved 2LPT density field does not capture completely the inherent cosmological dependence, in accordance with the results from \citet{he2018learning}.
    However, we expect that this sensitivity to the cosmology may be reduced by training over a range of cosmologies, thereby rendering our halo painting network more resilient to cosmological priors. We defer such an investigation to a future work.
    
    \subsection{Individual mass bins} \label{mass_bins_WGAN}
    
    To verify whether the network has properly encoded the halo mass information, we investigate the mapping learned for the different mass bins.
    We depict in Figs. \ref{fig:summary_statistics_individual_bins} and \ref{fig:bispectrum_individual_bins}, respectively, the corresponding power and bispectrum diagnostics for the four individual mass bins.
    Their respective predicted power and bispectra demonstrates the exceptional performance of our halo painting network across the range of halo masses involved.
    The above diagnostics show that the network is also capable of predicting the mass distribution of halos.
    
    \section{Summary and Conclusions} \label{conclusions_WGAN}

    We have presented a novel halo painting network\footnote{The source code repository, including a \textsc{Jupyter} notebook tutorial, is available at \url{https://github.com/doogesh/halo_painting}} for mapping 3D density fields to dark matter halo fields, inspired by an implementation of a recent variant of generative adversarial networks which employs the Wasserstein distance as a metric for training the network.
    Our network architecture encodes some of the recently proposed refinements to optimize its effectiveness and efficiency.
    The painting network employs residual Inception blocks, combining residual networks (ResNets) and Inception architecture, a choice which was heavily inspired by our physical notion of the true process.
    For improved training performance, we implement the gradient penalty method via the addition of a penalty term in the critic loss, as an alternative to the standard weight clipping, to enforce the Lipschitz-1 constraint on the critic.
    Our neural network, as a result of these upgrades, is not prone to training instability issues and does not exhibit any undesired behaviour.

\medskip    
    We train our halo painting network on 2LPT simulations to infer the relationship between the dark matter density field and the final halo distribution.
    We showcase the performance of our network in predicting the 3D halo distribution via a series of diagnostics, at the level of two- and three-point summary statistics, demonstrating that this mapping can be learned to a sufficiently high level of accuracy and that it reduces substantially the stochasticity over the Poisson shot noise.
    This performance is especially remarkable given that our neural network has only $\sim \mathcal{O}(10^4)$ trainable parameters, which is relatively few parameters in machine learning terms.
    In essence, our halo painting model allows us to rapidly generate simulations of halo distribution based on a non-linearly evolved density field within a fraction of a second on modern GPUs. 
    For instance, the network prediction for a $256^3$ simulation size requires roughly one second on the NVIDIA Quadro P6000.

\medskip
	A crucial aspect of our halo painting network, in a nutshell, lies in its capability to paint a halo distribution from a computationally cheap non-linear density field.
    This, as a result, provides a deterministic transformation to statistically populate the density field with highly non-linear structures such as halos.
    The halo painting network, therefore, bypasses the need to run full particle mesh simulations, thereby ensuring that detailed and high-resolution analyses of current and next-generation galaxy surveys, via the forward modelling approaches outlined below, are still feasible on regular computing clusters.
    Another interesting advantage of our approach is that our network can predict the 3D halo distribution for any arbitrary simulation box size due to the convolutional kernels being translationally invariant.
    With this method, a large simulation box does not require the tiling of smaller sub-elements, rendering our approach simple and elegant.
    
\medskip
    An immediate application of our halo painting network is that it can be employed for fast generation of mock halo catalogues and light cone production.
    This would be especially useful for the data analysis of upcoming large galaxy surveys of unprecedented sizes, such as Euclid and LSST.
    Another potential application of our network is that it can be utilized to fill in small-scale structure at a high resolution from low resolution large-scale simulations.
    It is also worth investigating whether we can further improve the performance of our painting model by training on the displacement field, rather than the density field, as carried out in \citet{he2018learning}.
    We defer such an investigation to a future undertaking.
    We also intend to explore other avenues, in terms of network architecture and training methodology, to render our halo painting model robust to different cosmologies, such that it could be optimized to constrain cosmological parameters via hierarchical Bayesian models, such as the \textsc{altair} (ALcock-Paczy\'nski consTrAIned Reconstruction) algorithm \citep{DKR2018altair}.

\medskip    
    The technology developed here is relevant to Bayesian forward modelling approaches for large-scale structure inference.
    For instance, it may be incorporated as part of the forward model in the \textsc{borg} (Bayesian Origin Reconstruction from Galaxies) \citep{jasche2013bayesian, jasche2015past, lavaux2016unmasking, jasche2018physical} framework  to generate sharp features due to redshift space distortions, such as Fingers of God effects.
    Since dark matter dynamics is well-posed, such a neural network may be employed as an emulator to transform an approximate large-scale model, such as Lagrangian Perturbation Theory (LPT), into a 3D halo distribution.
    Halo masses can subsequently be remapped to galaxy masses via another deterministic function, while encoding a stochastic selection, yielding the galaxy distribution of interest.
    
\medskip    
    Bayesian inference methods often require the adjoint gradient of the forward model in the sampling procedure, and as such, neural networks are perfectly suited for such tasks since they are, by their very nature, fully differentiable.
    The combination of statistical inference and machine learning is the plausible approach to accelerate high-resolution analyses of upcoming galaxy redshift surveys and provide statistically interpretable results, while maintaining the scientific rigour.

    \section*{Acknowledgements}
    We express our appreciation to the anonymous reviewer for their suggestions which helped to improve the overall quality of the manuscript.
	We thank Benjamin D. Wandelt for interesting discussions and for his comments on the paper.
    This work has been done within the activities of the Domaine d'Int\'er\^et Majeur (DIM) Astrophysique et Conditions d'Apparition de la Vie (ACAV), and received financial support from R\'egion Ile-de-France.
    DKR and GL acknowledge financial support from the ILP LABEX (under reference ANR-10-LABX-63) which is financed by French state funds managed by the ANR within the Investissements d'Avenir programme under reference ANR-11-IDEX-0004-02.
    This work was supported by the ANR BIG4 project, grant ANR-16-CE23-0002 of the French Agence Nationale  de  la  Recherche. This  work  was granted  access  to  the  HPC  resources  of  CINES  (Centre  Informatique National de l'Enseignement Sup\'erieur) under the allocation 2015047012 made by GENCI and  has  made  use  of  the  Horizon  cluster  hosted  by  the  Institut  d'Astrophysique  de  Paris  on  which  the  cosmological simulations  were  postprocessed.
    TC would like to thank NVIDIA for the Quadro P6000 used in this research.
    This work is done within the Aquila Consortium.\footnote{\url{https://aquila-consortium.org}}
    \bibliography{ML_references}
    \bibliographystyle{abbrvnat}

\end{document}